\documentclass[10pt, conference, letterpaper]{IEEEtran}

\usepackage{cite}
\usepackage[cmex10]{amsmath}
\usepackage{algorithmic}
\usepackage{array}
\usepackage{url}
\usepackage{footnote}
\usepackage{float}
\usepackage[utf8]{inputenc}

\usepackage[T1]{fontenc}
\usepackage[ruled,vlined]{algorithm2e}
\usepackage{ifthen}
\usepackage{graphicx}
\usepackage{comment}
\usepackage{amsthm}
\usepackage{amsmath}
\usepackage{amsmath,amssymb,amsfonts}
\usepackage{nicefrac}
\usepackage{algorithmic}
\usepackage{bm}
\usepackage{mathtools}
\usepackage{tikz}
\usetikzlibrary{calc,math}
\usepackage[framemethod=TikZ]{mdframed}
\usepackage{hyperref}
\hypersetup{colorlinks=true, unicode=true, linkcolor=[rgb]{0.10,0.05,0.67}, citecolor=[rgb]{0.10,0.05,0.67}, filecolor=[rgb]{0.10,0.05,0.67}, urlcolor=[rgb]{0.10,0.05,0.67}}

\newtheorem{theorem}{Theorem}

\newtheorem{remark}{Remark}

\newtheorem{example}{Example}

\newtheorem{assum}{Assumption}

\DeclareMathOperator*{\argmin}{arg\,min}

\definecolor{DarkGreen}{rgb}{0.1,0.5,0.1}
\definecolor{DarkRed}{rgb}{0.5,0.1,0.1}
\definecolor{DarkBlue}{rgb}{0.1,0.1,0.5}
\definecolor{DarkPurple}{rgb}{0.5,0.2,0.5}
\definecolor{DarkTurquoise}{rgb}{0.1,0.5,0.5}

\makeatletter
\renewcommand*{\eqref}[1]{%
  \hyperref[{#1}]{\textup{\tagform@{\ref*{#1}}}}%
}
\makeatother

\begin{document}

\title{Stream Iterative Distributed Coded Computing for\\
 Learning Applications in Heterogeneous Systems}
\author{
    \IEEEauthorblockN{Homa Esfahanizadeh$^{\dagger}$, Alejandro Cohen$^{\ddagger}$, and Muriel M\'edard$^\dagger$}
    \IEEEauthorblockA{
    $^\dagger$Massachusetts Institute of Technology (MIT), Cambridge, USA, Emails: \{homaesf, medard\}@mit.edu
    }
     \IEEEauthorblockA{
        $^\ddagger$Technion—Israel Institute of Technology, Haifa, Israel, Email: alecohen@technion.ac.il
    }
}

\maketitle

\begin{abstract}
To improve the utility of learning applications and render machine learning solutions feasible for complex applications, a substantial amount of heavy computations is needed. Thus, it is essential to delegate the computations among several workers, which brings up the major challenge of coping with delays and failures caused by the system's heterogeneity and uncertainties. In particular, minimizing the end-to-end job in-order execution delay, from arrival to delivery, is of great importance for real-world delay-sensitive applications. In this paper, for computation of each job iteration in a stochastic heterogeneous distributed system where the workers vary in their computing and communicating powers, we present a novel joint scheduling-coding framework that optimally split the coded computational load among the workers. This closes the gap between the workers' response time, and is critical to maximize the resource utilization. To further reduce the in-order execution delay, we also incorporate redundant computations in each iteration of a distributed computational job. Our simulation results demonstrate that the delay obtained using the proposed solution is dramatically lower than the uniform split which is oblivious to the system's heterogeneity and, in fact, is very close to an ideal lower bound just by introducing a small percentage of redundant computations.
\end{abstract}

\begin{IEEEkeywords}
distributed systems, coded computation, heterogeneous, straggler, scheduling.
\end{IEEEkeywords}

\section{Introduction}

With the emerging advances in the storage and computing resources and thanks to the data accessibility, the demand for iterative computational algorithms in machine learning (ML) applications, adaptive control algorithms, numerical optimizations, etc, has grown significantly over the last decade. Although, simple iterative algorithms such as training a small ML  model with a moderate-size data can be locally performed over one server, distributing the computations is essential for many advanced iterative algorithms that require massive amounts of data and very high processing power in order to be trained and utilized in practice, e.g., automatic translation, image localization, and playing games.

\begin{figure}
    \centering
    \vspace{0.15cm}
    \includegraphics[width = 0.95 \columnwidth]{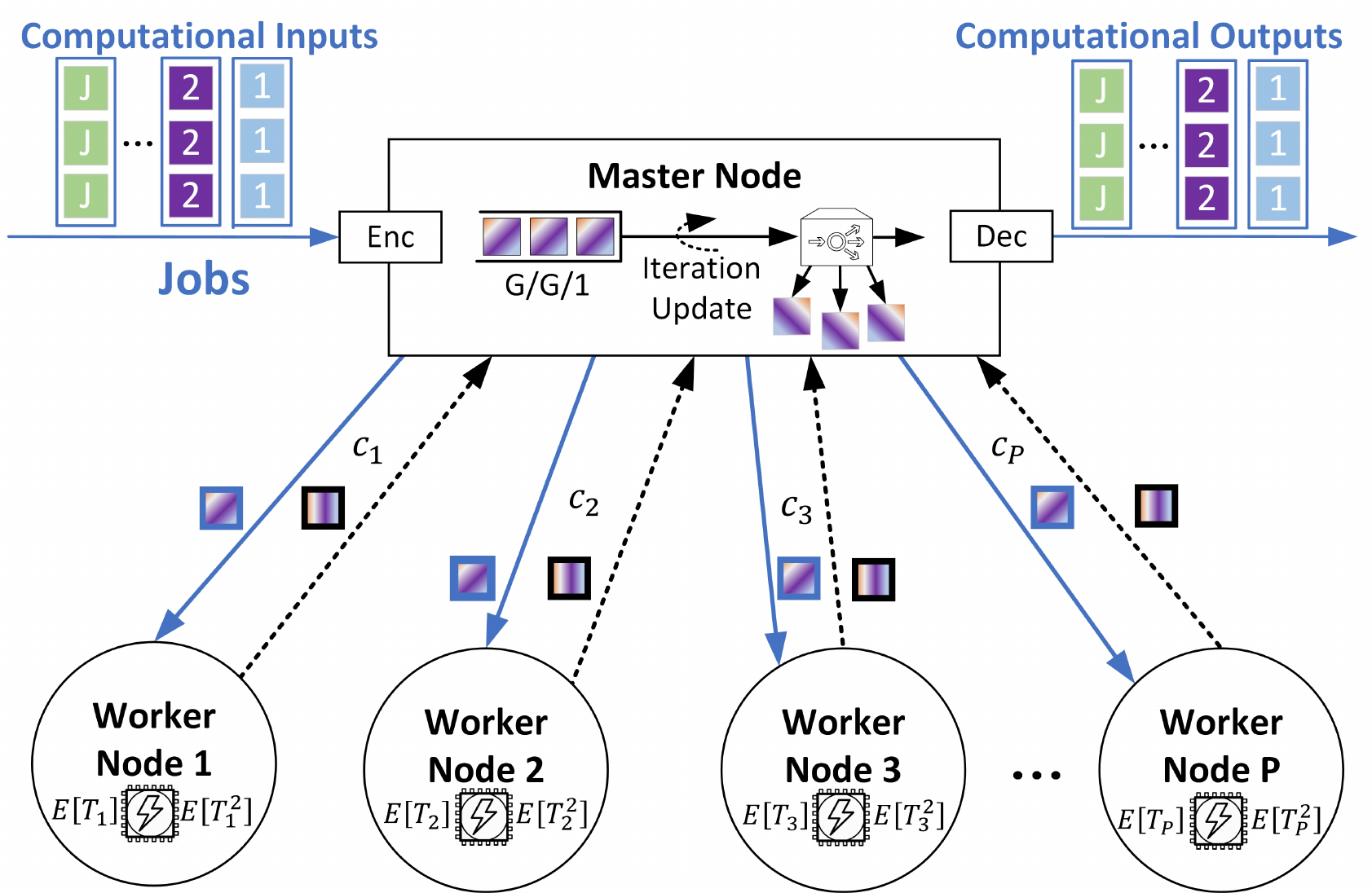}
    \caption{System model for the stream iterative distributed computation problem. The parameters are: $E[T_p]$ and $E[T_{p}^2]$ for the first and second moment of the average computation time, respectively. $c_p$ is for the communication time per iteration. The master node is modeled with a G/G/1 queue.\vspace{-0.5cm}}
    \label{fig:system_model}
\end{figure}

Motivated by the low latency demands of modern computational applications, e.g., in smart cities and autonomous driving, one major goal of a distributed computation solution is to reduce the average in-order execution delay, which is defined as the average time a computational job spends in the system (from arrival to delivery). This goal is even more critical for distributed computation of iterative jobs since in such setting the next iteration cannot start until the previous iteration is finished. Yet, even recent  state-of-the-art  implementations of distributed ML models have not reported to  achieve expected scalability (notable speed up by increasing the number of machines)  because of not efficiently utilizing all resources\cite{XINXingML}. Fig.~\ref{fig:system_model} depicts our problem setting for distributed computation of a stream of iterative computational jobs, and it consists of a master nodes and a heterogeneous cluster of worker nodes. The master node splits the computational load of each iteration of a job among the workers, aggregate the results, and move to the next iteration or first iteration of the next job in its queue. 

One important consideration in a distributed computing system is to deal with delays and failures caused by slow workers, called stragglers \cite{dean2013tail,joshi2017efficient}. In fact, we consider a more general and realistic setting where the workers are heterogeneous in their computing and communicating capabilities \cite{Reisizadeh,Avestimehr}. In an ideal solution, the master node does the task assignment such that it takes the maximum benefit from all resources rather than discarding or being restricted by the slow ones. There have been two mechanisms for dealing with the system heterogeneity and uncertainties: The first one is to introduce redundant computations into the computational load that is split among the workers, known as distributed coded computations, e.g., \cite{sheth2018application,d2020gasp,wahabfederated}, and the second one is to use non-uniform load balancing among the workers considering their various capabilities \cite{Reisizadeh,cohen2021stream}. In what follows, the previous works in these two directions are briefly described.

The distributed coded computation has recently attracted significant attention as it provides redundant computational load per job (or job iteration) in order to compensate for straggling effects in advance, e.g., \cite{yu2017polynomialn,raviv2020gradient}. Since the assigned computational load is more than necessary, the master node does not need to wait for each worker to finish all its assignment to move on to the next job (or iteration). In fact, as soon as the master node collects enough results to resolve the job (or job iteration), it requests the workers to drop the remaining part of their assignment related to the finished job (or job iteration); and such a process is called \textit{purging}. Next, we review some recent examples of distributed coded computation, under the category of computational jobs that they target.

For the gradient descent algorithm which is an essential iterative algorithm used in many learning applications \cite{karakus2019redundancy}, various coding schemes have been considered to introduce the redundant computations, e.g., maximum-distance-separable (MDS) codes \cite{raviv2020gradient,dutta2019short,tandon2017gradient}, Reed-Solomon codes \cite{halbawi2018improving}, LDPC codes \cite{maity2019robust}, and neural-network-based codes \cite{kosaian2018learning}. For matrix-matrix and matrix-vector multiplication which are major building blocks for many learning applications, MDS codes, Short-Dot codes, among many others \cite{8002642,dutta2019short,8437852,dutta2019short,lee2017high,suh2017matrix,baharav2018straggler,yu2017polynomialn,Dutta2020,mallick2019fast}, have been considered or introduced. The relation between communication redundancy (the traffic load that need to be communicated to the workers) and computation redundancy (the redundant computational load at the workers) have been incorporated into the design, e.g., see \cite{dutta2019short,8437852,Maddah-Ali2018CommComp,ye2018communication}. Finally, for data shuffling, convolution, and fast Fourier transform, distributed coded computation methods have been proposed, e.g., \cite{song2019pliable,attia2019near,dutta2017coded,yang2016fault,jeong2018masterless,yu2017coded}. 

Majority of previous works on distributed coded computation topic do not consider heterogeneity of the workers into the design, and they introduce redundant computations to deal with possible stragglers. This view although increases the reliability is not sufficient for applications where having a low in-order execution delay is critical. Motivated by this necessity, another orthogonal mechanism is needed for coping with the system heterogeneity via a non-uniform load split. Recently, \cite{Reisizadeh,Avestimehr} proposed coding methods that asymptotically obtain the optimal solutions in a heterogeneous system with specific stochastic behaviour. More recently, the optimal load balancing in conjunction with the distributed coded computation has been proposed for non-iterative computational jobs with general stochastic description to minimize the average end-to-end job delivery delay \cite{cohen2021stream}. However, the results of \cite{cohen2021stream} are not extendable to iterative jobs due to the dependencies of the iterations of a job to each other.

In this paper, we propose a new joint scheduling-coding solution for distributed coded computation of a delay-sensitive stream of iterative jobs over a heterogeneous set of workers, where we efficiently exploit both mechanisms, i.e., redundant computations and load balancing, to reduce the in-order execution delay. To this purpose, we borrow tools and techniques from the queuing theory, and we in fact model the whole system as a G/G/1 queuing model \cite{Klein1975}. To encompass the most general case, the arrival of incoming iterative jobs at the master node is assumed to have a general distribution, and the total iterations of a job, distributed and computed at the workers per iteration, is considered as a service time with a general distribution, which will be described in this paper. 

In particular, we propose an optimal load balancing that makes the distribution of the time it takes for various workers to finish their assignment as close to each other as possible, and show this results in the minimum in-order job execution delay. To this end, we consider a distribution distance measure that only depends on the communication rate and the first and second moments of the service time at each worker, which can be provided for the master node by workers' declaration or be estimated during the run-time. We also investigate the effect of coding parameters on the overall behaviour of the distributed system and propose optimizing them to further reduce the execution delay. 

Our simulation results demonstrate that our proposed solution obtains almost the same in-order execution delay as an ideal genie-aided scheme by introducing an incremental amount of redundant computations. The proposed solution is general in the sense that it can be used with many existing coding schemes and probabilistic models. We note that in our system model, we assume an entire iteration of a job (including linear and non-linear operations) is split among the workers, and thus, our model is different from \cite{8437852} where each iteration is divided to several linear and non-linear steps and the linear parts are split among the workers in sequential steps. In particular, according to our simulation results for distributed coded gradient descent, the proposed solution outperforms the uniform solution considered, e.g., in \cite{dutta2019short,tandon2017gradient}; it reduces the job in-order execution delay by a factor of more than two and a half and can approach the lower bound by just incorporating  less than $10\%$ redundant computations.

The rest of this paper is organized as follows: In section~II, we describe the system model and the problem setting. In Section~III, we propose our optimal
load balancing solution
for stream distributed coded computation of iterative jobs. In Section~IV, we perform an analytical study of the system and also propose a lower-bound on the job execution delay. In Section~V, we investigate the effect of coding parameters on the optimal load balancing and correspondingly on the job execution delay. Sections~VI and VII dedicate to simulation results and conclusions, respectively. Finally, an example of coding schemes for distributed gradient descent algorithm is presented in the appendix. 

\section{System Model and Problem Setting}

The system model is composed of a master node and several worker nodes, see Fig.~\ref{fig:system_model}. The master node sequentially receives iterative computational jobs of the same size and processes them, in order, by delegating the computations to the workers. The inter-arrival time of the jobs are independent from each other, and is modeled with random variable $T_a$ with given first moment $E[T_a]$ and second moment $E[T_a^2]$. The master node starts with first iteration of the first job in the queue and assigns appropriate portion of the computations to each worker in a set of workers $\mathcal{P}$, where $|\mathcal{P}|=P$. Each worker independently performs pre-defined computational operations on the assigned tasks and sends the results back to the master node. The master node aggregates the results to produce the result of one iteration and moves forward to the next iteration or the first iteration of the next job in the queue. 

Let $\mathcal{I}$ be the number of iterations per job. Each iteration of a job is divided into several smaller computational tasks, and the master node requires to obtain $K$ completed task results per job iteration to identify the result of that iteration. However, to compensate for stragglers, the master node assigns $K\Omega$ tasks, where $\Omega\geq1$, to workers per job iteration. The parameters $K$ and $\Omega$ are called number of critical tasks and redundancy ratio, respectively. Each worker sends the result to the master node as soon as one task is finished and does not wait for all the assigned tasks to finish. We denote as $\kappa_p$ the number of tasks per job iteration that is assigned to the $p$-th worker, thus
\begin{equation*}
    \sum_{p=1}^{P}\kappa_p=K\Omega.\vspace{-0.2cm}
\end{equation*}
 
We denote with $T_{p}$ and $T_{p,\kappa_p}$ the time it takes for $p$-th worker to finish one task and the whole assignment of one job iteration (communication with the master node and performing $\kappa_p$ sequential tasks), respectively. Thus, we model $T_{p,\kappa_p}$ to be summation of $\kappa_p$ independent random variables, each with the same distribution as $T_p$, plus a fixed shift $c_p$ that denotes the communication time between master node and the $p$-th worker per job iteration,\footnote{A constant communication delay assumption is suited for applications where workers transmit their local updates, e.g., updated gradients of an ML model, whose sizes are independent of the number of assigned tasks.}\vspace{-0.5cm}
\begin{equation}\label{equ:def_T_p_kappa_p}
T_{p,\kappa_p}=c_pI_{\kappa_p}+\overbrace{T_p+\dots+T_p}^{\kappa_p\text{ times}}.
\end{equation}
Here $I_{\kappa_p}$ is the indicator function that takes value $1$ if $\kappa_p>0$, and $0$ otherwise, and it appears since when no task is assigned to a worker, it does not need to communicate with the master node. As an example, in distributed gradient descent problem, $c_p$ stands for the time it takes to transfer the ML model updates at each job iteration. 

Next, let $T_\text{itr}$ be the time it takes for the system to finish one iteration of a job. According to the system model,
\begin{equation}
\label{equ_max}
    T_\text{itr}\leq\max\{T_{1,\kappa_1},\dots,T_{P,\kappa_P}\}.
\end{equation}
The equality in (\ref{equ_max}) holds when the redundant computations are not removed from the system once the iteration is finished, i.e., no purging occurs.
Finally, the service time of the system denoted with $T_s$ is summation of $\mathcal{I}$ iid random variables, each with the same distribution as $T_\text{itr}$. An example of coding scheme for distributed gradient descent algorithm that can be combined with the presented joint scheduling-coding framework in this paper is given in the appendix. Table~\ref{tab:parameters} shows a list of frequently visited parameters throughout this paper.

\begin{table}
        \caption{Table of frequently visited parameters.\vspace{-0.2cm}}
    \centering
    \begin{tabular}{|l|l|}
    \hline
    Parameter & Definition\\
    \hline 
    \hline 
    $T_a$&iterative job inter-arrival time\\
    \hline
    $T_s$&iterative job service time\\ 
    \hline
    $T_\text{itr}$&iteration service time\\ 
    \hline
    $T_{p}$, $T_{p,\kappa_p}$&task and assignment service time for $p$-th worker\\
    \hline
    $\mathcal{I}$&number of iterations\\
    \hline
    $\lambda$ & average job arrival rate\\
    \hline
    $\mathcal{P}$ and $P$& set of available workers and its cardinality\\
    \hline
    $\kappa_p$ & number of tasks per job iteration assigned to $p$-th worker\\
    \hline
    $K$&number of critical tasks per job iteration\\
    \hline
    $\Omega$&redundancy ratio\\
    \hline
    \end{tabular}\vspace{-0.5cm}
    \label{tab:parameters}
\end{table}




\section{Load Split Optimizations}
In this section, we first show that the solution with maximum utilization results in the minimum service time. Next, we propose an optimal load split to reduce the job delay by maximizing the utilization.

\subsection{Compensating for Heterogeneity by Load Balancing}

Let the random variable $r_p(t)$, $p\in\{1,\dots,P\}$, denote the portion of one job iteration that the $p$-th worker performs at time step $t$. We also define the binary state function $u_p(t)$, $p\in\{1,\dots,P\}$, such that its value $1$ shows the worker is active at time step $t$ and its value $0$ shows the worker is idle at time step $t$. An iteration of a job finishes as soon as the combined portion of computational results collected from all workers related to one job iteration exceeds value one, i.e., the minimum value $T_\text{itr}$ such that
\begin{equation*}
    \sum_{p=1}^{P}\sum_{t_\circ}^{t_\circ+T_\text{itr}}r_p(t)u_p(t)\geq1,
\end{equation*}
where $t_\circ$ represents start time of the iteration. This is because as soon as the inequality is satisfied, the master node is able to resolve the iteration and assigns to the workers the tasks related to next iteration of this job, or first iteration of the next job in the queue. 

We note the workers' state function $\{u_1(t),\dots,u_P(t)\}$ are the only parameters that depend on the distributed computation solution and thus can be optimized. For now, we assume the designer has full control over the value of state functions via the load balancing (optimizing the number of tasks assigned to each worker); however, in practice due to the stochastic behaviour of the system, communication delays, and fixed-point issues, one cannot obtain the exact desired choices for the state functions. We will extensively comment on this matter later in the paper. The next theorem shows that the minimum delay of a job iteration is obtained when all workers finish their assignments at the same time.
\begin{theorem}\label{theorem:utilization}
The minimum execution time $T_\text{itr}$ for one iteration of a job  that starts at time step $t$ is obtained when
$u_p(t)=1$ for every $t\in\{t_\circ,\dots,t_\circ+T_\text{itr}\}$ and $p\in\{1,\dots,P\}$.
\end{theorem}
\begin{proof}
We, in fact, need to identify the solution for the following optimization problem:
\begin{equation*}
\begin{split}
    &\argmin_{u_1(t),\dots,u_P(t)}T_\text{itr},\quad\text{s.t.}\quad
    \sum_{p=1}^{P}\sum_{t_\circ}^{t_\circ+T_\text{itr}}r_p(t)u_p(t)\geq1.
\end{split}
\end{equation*}
Since $T_\text{itr}$ is decreasing with respect to $u_p(t)\in\{0,1\}$, the optimal solution is obtained when $u_p(t)=1$, for every $t\in\{t_\circ,\dots,t_\circ+T_\text{itr}\}$ and $p\in\{1,\dots,P\}$, and this is regardless of nature of the random variable $r_p(t)$.
\end{proof}

We highlight that the time it takes for each worker to finish its assignment, which characterizes $u_p(t)$ for $p\in\{1,\dots,P\}$, has a stochastic behavior. Therefore, our load balancing optimization is to assign appropriate portion of the computational load of one job iteration to each worker such that the distributions of the time it takes for each worker to finish its assignment are as close as possible.

\subsection{Optimal Load Split Solution}

We define $T_\text{ref}$ as a reference random variable with set of parameters represented by $\Theta$. Then, we optimize the load split such that the assignment service time $T_{p,\kappa_p}$ at each worker $p\in\{1,\dots,P\}$ has a close distribution to distribution of $T_\text{ref}$. For measuring the distance between the two distributions, denoted with $D(T_{p,\kappa_p}||T_\text{ref})$, one can use variety of distance metrics, e.g., Kullback–Leibler (KL) divergence, Lévy–Prokhorov metric, etc \cite{mackay_2019}. Then, the optimization problem can be described as follows,\footnote{We relax the optimization problem and let $\kappa_p$ take real values. In practice, we choose the closest integer to the optimal values such that $\sum_{p=1}^{P}\kappa_p{=}K\Omega$.}\vspace{-0.2cm}
\begin{equation}\label{opt:1}
\begin{split}
    &\min_{\Theta,\kappa_1,\dots,\kappa_P}\sum_{p=1}^{P}D(T_{p,\kappa_p}||T_\text{ref}),\quad\kappa_p\geq0,\;\sum_{p=1}^{P}\kappa_p=K\Omega.
\end{split}
\end{equation}
In this paper, we use the following distance measure thanks to its simplicity and the availability of its required information:
\begin{equation}\label{equ:distance}
     D(T_{p,\kappa_p}||T_\text{ref})=(E[T_{p,\kappa_p}]+\gamma E[T_{p,\kappa_p}^2]-\theta)^2,
\end{equation}
where $\theta\triangleq E[T_\text{ref}]+\gamma E[T_\text{ref}^2]\geq0$.
\begin{remark}
The notion of distribution distance in \eqref{equ:distance} is particularly helpful when the exact distributions for the assignment service time at the workers are not known, and instead, their moments are obtained via declaration and/or feedback-based estimation. By extending this measure to include the higher moments, we obtain the square of Euclidean distance between the moment generating functions of the two distributions. We note that \eqref{equ:distance} is a heuristic measure of distribution distance, and we empirically show in the simulation results that an optimal solution obtained via this distance measure achieves on average a close job execution delay to a theoretically-driven lower bound.
\end{remark}
The parameter $\gamma>0$ adjusts the relative importance of the first moment and the second moment on the distance measure. Next, we identify the first moment and second moment of $T_{p,\kappa_p}$, $p\in\{1,\dots,P\}$. According to \eqref{equ:def_T_p_kappa_p},\vspace{-0.1cm}
\begin{align*}
    E[T_{p,\kappa_p}]&=c_pI_{\kappa_p}+\kappa_pE[T_p],\\
    E[T_{p,\kappa_p}^2]&=c_p^2I_{\kappa_p}+2\kappa_p c_pE[T_p]+\kappa_pE[T_p^2]\\
    &+\kappa_p(\kappa_p-1)E[T_p]^2.
\end{align*}
Thus, we rewrite the optimization problem \eqref{opt:1} as follows:\vspace{-0.4cm}

\begin{equation}\label{eq:opt2}
\begin{split}
    \min_{\theta,\kappa_1,\dots,\kappa_P}&\sum_{p=1}^P(a_pI_{\kappa_p}+b_p\kappa_p+\gamma m_p^2\kappa_p^2-\theta)^2,\\
    &\text{s.t. }\quad\kappa_p\geq0,\quad\theta\geq0,\quad\sum_{p=1}^{P}\kappa_p=K\Omega,
\end{split}
\end{equation}
where 
\begin{equation*}
\begin{split}
    m_p&\triangleq E[T_p],\quad\sigma_p^2\triangleq{E[T_p^2]-E[T_p]^2},\\
    a_p&\triangleq c_p+\gamma c_p^2,\\
    b_p&\triangleq m_p+2\gamma c_p m_p+\gamma\sigma_p^2.
\end{split}
\end{equation*}

Next, we identify the solution for problem \eqref{eq:opt2}:
\begin{theorem}\label{the:opt_sol}
The optimal load split among the workers is
\begin{equation*}
    \kappa_p=\frac{b_p}{2\gamma m_p^2}{\left(-1+\sqrt{1+\frac{4\gamma m_p^2(\theta-a_p)^+}{b_p^2}}\right)},
\end{equation*}
where $(\alpha)^+=\max\{\alpha,0\}$ and $\theta$ is set to make $\sum_{p=1}^{P}\kappa_p=K\Omega$.
\end{theorem}
\begin{proof}
Without loss of generality, let assume the set of active workers is $\mathcal{P}^a\subseteq\{1,\dots,P\}$. Thus, for every $p\in\mathcal{P}^a$, $\kappa_p>0$; and for every $p\notin\mathcal{P}^a$, $\kappa_p=0$. Therefore, the minimum possible value for the objective function is $(P-|\mathcal{P}^a|)\theta^2$, and is obtained when
\begin{equation*}
    a_pI_{\kappa_p}+b_p\kappa_p+\gamma m_p^2\kappa_p^2-\theta=0,
\end{equation*}
for every $p\in\mathcal{P}^a$, i.e.,
\begin{equation*}
    \kappa_p=\frac{b_p}{2\gamma m_p^2}{\left(-1+\sqrt{1+\frac{4\gamma m_p^2(\theta-a_p)}{b_p^2}}\right)}.
\end{equation*}
Since $\kappa_p$ must be positive for the active workers, it requires $a_p<\theta$ for every $p\in\mathcal{P}^a$. As a result, $\theta$ determines the set of active workers as follows:
\begin{equation*}
    \mathcal{P}^a=\{p\in\mathcal{P}:c_p+\gamma c_p^2<\theta\}.
\end{equation*}
The value of $\theta>0$ is set such that $\sum_{p=1}^{P}\kappa_p=K\Omega$. We note that $\sum_{p=1}^P{\kappa_p}$ is strictly increasing with respect to $\theta$ (by increasing $\theta$, the value of $\kappa_p$ increases for already active workers and also more workers may get activated). Therefore, such unique $\theta$ can always be found using a binary search.
\end{proof}
\begin{example}
Let $c_p=0$ and $T_p\sim\text{Exp}(\mu_p)$ for every $p\in\{1,\dots,P\}$. In this setting, $a_p=0$ for every $p\in\mathcal{P}$ and thus all workers are active. Besides, $m_p=\sigma_p=1/\mu_p$ and $b_p=(\mu_p+\gamma)/\mu_p^2$. As a result,
\begin{equation*}
\begin{split}
    \kappa_p&=\frac{b_p}{2\gamma m_p^2}{\left(-1+\sqrt{1+\frac{4\gamma m_p^2\theta}{b_p^2}}\right)}\\
    &=\frac{(\mu_p+\gamma)}{2\gamma}\left(-1+\sqrt{1+\frac{4\gamma\mu_p^2 \theta}{(\mu_p+\gamma)^2}}\right).
\end{split}
\end{equation*}
\end{example}

In the next section, we identify the analytical expression for the average job execution delay, investigate the rate stability of the queue at the master node, and also propose a lower bound on the delay obtained by the optimal load balancing solution.

\section{Analytical Investigations}

We first remind that according to the system model, $E[T_s]=\mathcal{I}E[T_\text{itr}]$ and 
$T_\text{itr}=\max\{T_{1,\kappa_1},\dots,T_{P,\kappa_P}\}.$ Let represent the distribution of the time it takes for each worker to finish its assignment (related to a job iteration) as $F_p(t)=P[T_{p,\kappa_p}\leq t]$ for every $p\in\{1,\dots,P\}$ (CDF of the $T_{p,\kappa_p}$), and let $f_p(t)$ be the derivative of  $F_p(t)$ with respect to $t$ (PDF of  $T_{p,\kappa_p}$). Thus,
\begin{equation*}
\begin{split}
    &F_\text{itr}(t)=P[T_\text{itr}\leq t]=\prod_{p\in\mathcal{P}^a}F_p(t),\\
    &f_\text{itr}(t)=\frac{\mathrm{d}F_\text{itr}(t)}{\mathrm{d}t}=\sum_{p'\in\mathcal{P}^a}f_{p'}(t)\hspace{-0.2cm}\prod_{p\in\mathcal{P}^a\setminus \{p'\}}\hspace{-0.2cm}F_p(t).
\end{split}
\end{equation*}

\subsection{Rate Stability Analysis}
Here, we consider the rate stability for the G/G/1 queue at the master node, which is described by the following constraint,
\begin{equation*}
    E[T_s]<E[T_a].
\end{equation*}
Given pdf of $T_\text{itr}$ i.e., $f_\text{itr}(t)$, we can identify $E[T_\text{itr}]$, $E[T_s]=\mathcal{I}E[T_\text{itr}]$, and then determine if the system is stable or not for a given load split choice. Thus, given a set of workers and the optimal split, if the system is not stable, we need to add more workers and repeat the load balancing optimization until we find an stable solution.

\begin{remark}
If an optimal split with parameter $\theta$ does not provide rate stability for a set of workers, adding a worker $p$ with $a_p\geq\theta$ does not make the system stable as it will not be utilized in the optimal load split.
\end{remark}

\begin{remark}
The presented stability analysis is for the extreme case where the redundant computations are not removed from the system when the iteration is finished. Thus, if satisfied, it also satisfies the rate stability when purging occurs.
\end{remark}

\subsection{Execution Delay Analysis}

For G/G/1 queing model, the average response time is approximated by Kingman's formula\cite{marchal1976approximate}:
\begin{equation}\label{eq:Kingmansformula}
    D\approx E[T_s]\left(1+\frac{\rho}{1-\rho}\frac{c_a^2+c_s^2}{2}
    \right),
\end{equation}
which is known to be generally very accurate, especially for a system operating close to saturation. Here $\rho=E[T_s]/E[T_a]$, $c_a^2=(E[T_a^2]-E[T_a]^2)/E[T_a]^2$, and $c_s^2=(E[T_s^2]-E[T_s]^2)/E[T_s]^2$. When the job arrival is modeled with a Poisson distribution with parameter $\lambda$ (i.e., M/G/1 queuing model), the average job delay of the system is given more precisely by Pollaczek-Khinchin formula \cite[Chapter 5]{gallager2013stochastic}:
\begin{equation}\label{eq:PK}
    D=E[T_s]+\frac{\lambda E[T_s^2]}{2(1-\lambda E[T_s])}.
\end{equation}

Here, we assume all $K\Omega$ tasks are necessary for finishing an iteration of a job. Thus, the derived expression for the average job delay is an upper bound for the real-world value where only $K$ tasks out of $K\Omega$ tasks are needed to finish the iteration, and the redundant tasks are removed from the system once the iteration is finished. The derived expression is precise for the case when $\Omega=1$. Given the PDF of $T_\text{itr}$, we can identify $E[T_\text{itr}]$ and $E[T_\text{itr}^2]$. Then,
\begin{equation}
\begin{split}
    &E[T_s]=\mathcal{I} E[T_\text{itr}],\\
    &E[T_s^2]=\mathcal{I} E[T_\text{itr}^2]+\mathcal{I}(\mathcal{I}-1)E[T_\text{itr}]^2.
\end{split}
\end{equation}
By substituting these terms in \eqref{eq:Kingmansformula} or \eqref{eq:PK}, the theoretical value for the average job execution delay of the system is obtained for the no-purging case.

\subsection{Theoretical Lower Bound for Job Execution Delay}

Let consider computational load of each job iteration is partitioned into several tasks and $K$ of them are critical to resolve the job iteration. We remind that the task service rate of the $p$-th worker is $1/E[T_p]$ and its communication delay per iteration is $c_p$. Thus, a lower bound on the system delay is obtained by approximating the whole system with just one worker whose task service rate is summation of the task service rates of all workers and its delay is the average delay of the workers. This is because such an approximation models the maximum utilization (the modeled worker does not need to wait for results of other workers to start a new iteration). The following average job delay is obtained for this approximated system:
\begin{equation}\label{eq:LB}
    D_\text{L}=\mathcal{I}\left(\frac{K}{\sum_{p=1}^{P}\frac{1}{E[T_p]}}+\frac{1}{P}\sum_{p=1}^P{c_p}\right),
\end{equation}
which acts as a lower bound on the job execution delay of the real system. We remind that the system requires to service $K$ tasks per iteration and each job consists of $\mathcal{I}$ iterations.

\begin{example}[Distributed coded gradient descent algorithm] In the appendix, we review the distributed gradient descent algorithm as an example of iterative algorithms that can be incorporated into our framework. The distributed gradient descent suggests performing the gradient computations in a distributed fashion over smaller chunks of the dataset distributed among the workers. For this purpose, the dataset with size $n$ samples is partitioned into $m$ disjoint chunks each with size $n/m$. Each task is performing the SGD individually on $d$ disjoint chunks of the dataset and sending a linear combination of the results to the master node. The master node is capable of obtaining the full gradient update from any $K$ task results, and then, it completes the iteration by performing the weight update of the model. The complexity of SGD algorithm is approximately linear with size of the samples it performs over, i.e., $n/m$, and each task consists of running $d$ individual gradient descent algorithms, one per each assigned chuck of dataset. Therefore, the complexity of $d$ tasks is $C\approx d\alpha\frac{n}{m}$. In this example, we assume, $n=554,400$ samples, $m=100$, $d=51$, $\alpha=10$, and therefore, $C\approx2,827,440$. Besides, we assume $K=50$ and $\Omega=1.1$. 

Next we describe the parameters of the heterogeneous system. Let the job arrival be modeled as a Poisson random process with rate $\lambda=0.01$ and the number of workers be $P=5$, each with exponential task service time $T_p\sim\text{Exp}(\mu_p/C)$ and communication delay $c_p$. These parameters are listed below:
\begin{table}[H]
    \centering
    \begin{tabular}{|c|ccccc|}
       \hline $p$&$1$&$2$&$3$&$4$&$5$\\
       \hline
        $\mu_p$&5.29E7&7.26E7&3.10E7&1.37E7&6.03E7\\
        \hline
        $c_p$&0.0481&0.0562&0.0817&0.0509&0.0893\\
        \hline
    \end{tabular}
\end{table}
\noindent We assume the number of iterations is $\mathcal{I}=50$. For this example, the average job execution delay computed over $J=1000$ simulated arrived jobs is $47.93$ seconds for the proposed optimal solution and $\gamma=1$, whereas it is $129.96$ seconds for the uniform solution as assumed in e.g., \cite{dutta2019short,tandon2017gradient,raviv2018gradient,halbawi2018improving,raviv2020gradient}. The theoretical lower bound for the average job execution delay is $42.04$ seconds, see \eqref{eq:LB}.
\end{example}

\section{Optimization of Code Parameters}

In this section, we study the role of code parameter optimization for reducing the job execution delay in the distributed coded computation problem. We highlight that if the communication delay is negligible and the computational load can be divided into any arbitrary number of tasks ($\kappa_p$ can take any real values), the optimal solution corresponds to matching all distributions together. However, when the communication delay is not negligible and the quantization error occurs, there exist a mismatch error that is quantified as follows:
\begin{equation}\label{eq:mismatch}
    \textit{mismatch}=\textit{var}(\{E[T_{p,\kappa_p}]+\gamma E[T_{p,\kappa_p}^2]\}_{p\in\mathcal{P}}),
\end{equation}
where $\textit{var}(.)$ computes the numerical variance of a sequence. In this section, we propose minimizing the mismatch by an optimal choice of the code parameters.

Here, the complexity of one task is denoted with parameter $C$. Thus, the set $\{K,C,\Omega\}$ encompasses the code parameters, and we denote with $\textit{Codes}$ the set of all possible sets of code parameters (all possible codes). Let $U^*_p$ be the time it takes for the $p$-th worker to finish one task with complexity $1$, and the master node has access to $E[U_p]$ and $E[U_p^2]$ through the workers' declaration and/ or feedback-based estimations during the execution.
\begin{assum}\label{assum:mother_dist} The distribution of the time it takes for the $p$-th worker to finish one task with complexity $C$ is
$$P[T_p\leq t]=P[U_p\leq \frac{t}C].$$  
\end{assum}
\noindent As a result, first moment and second moment of the time it takes for the $p$-th worker to finish one task is $E[T_p]=CE[U_p]$ and $E[T_p^2]=C^2E[U_p^2]$, respectively. This assumption is in accordance with model of mother runtime distribution in \cite{lee2017speeding,baharav2018straggler}. 

Algorithm~\ref{alg1} shows our procedure to find the optimal code parameters that result in the minimum job execution delay using in conjunction with the proposed optimal load split in this paper. We note that the set of possible code parameters has either a computationally-affordable size or can be managed or pruned by the designer. Thus, the complexity of the algorithm is affordable.

\begin{algorithm}
\SetAlgoLined
\KwResult{Optimal code parameters and load split}
\textbf{Initialization:} $\nu^*=\infty$\\
\For{$\{K,C,\Omega\}\in \textit{Codes}$}{\text{ }\\
    Using Assumption~\ref{assum:mother_dist}, find $E[T_p]$ and $E[T_p^2]$\\
    Find  $\{\kappa_1,\dots,\kappa_P\}$ using Theorem~\ref{the:opt_sol}\\
    Quantize $\{\kappa_1,\dots,\kappa_P\}$ to have integer values\\
    Identify the \textit{mismatch} using \eqref{eq:mismatch}\\
    \If{$\textit{mismatch}<\nu^*$}{\text{ }\\
        $\{K,C,\Omega\}^*:=\{K,C,\Omega\}$\\
        $\{\kappa_1,\dots,\kappa_P\}^*:=\{\kappa_1,\dots,\kappa_P\}$\\
        $\nu^*:=\textit{mismatch}$
    }
 }
 \caption{Optimization of Code Parameters and Load Split in Stream Distributed Coded Computation}
 \label{alg1}
\end{algorithm}

In this Algorithm, each choice for the set of code parameters is considered individually, and its \textit{mismatch} parameter corresponding to the optimal load split, presented in Theorem~\ref{the:opt_sol}, is identified using \eqref{eq:mismatch}. Then, the optimal set of code parameters and its corresponding optimal load split is reported as the final results.

\section{Simulation Results}

In this section, we evaluate performance of the proposed joint scheduling-coding framework for stream distributed coded computation of iterative jobs. First, we compare the optimal solution with a baseline solution that is oblivious to the system's heterogeneity in a streaming scenario. Next, we investigate the in-order job execution delay of the proposed solution with respect to the redundancy ratio and show how quickly it approaches to a theoretically-driven lower bound. Finally, we study the effect of coding parameters on performance of the optimal solution. 

We first describe the system parameters: The job arrival is modeled with a Poisson distribution with rate $\lambda=0.01$, and thus $E[T_a]=100$ and $E[T_a^2]=20,000$. We consider $\gamma=1$, and we assume each job consists of $\mathcal{I}=10$ iterations. We also assume $T_p\sim\text{Exp}(\mu_p/C)$, where $C$ is the number of operations per task, and thus, $T_{p,\kappa_p}$ is a shifted gamma distribution with shift $c_p$, shape $\kappa_p$, and scale $C/\mu_p$ when $\kappa_p>0$. In this section, we assume $Z\triangleq KC$ is fixed, and thus $K$ is inversely proportional to $C$.\footnote{Given the computational coding scheme, one can change this relation accordingly and repeat this set of simulations.} By optimal, resp., uniform, split we mean $\{\kappa_1,\dots,\kappa_p\}$ are set to be the closest integers to result of Theorem~\ref{the:opt_sol}, resp., $K\Omega/P$, such that they sum up to $K\Omega$.

\subsection{Streaming Computations: Optimal vs. Uniform}

\begin{figure}
    \centering
    \begin{tabular}{cc}
     \includegraphics[width = 0.435\columnwidth]{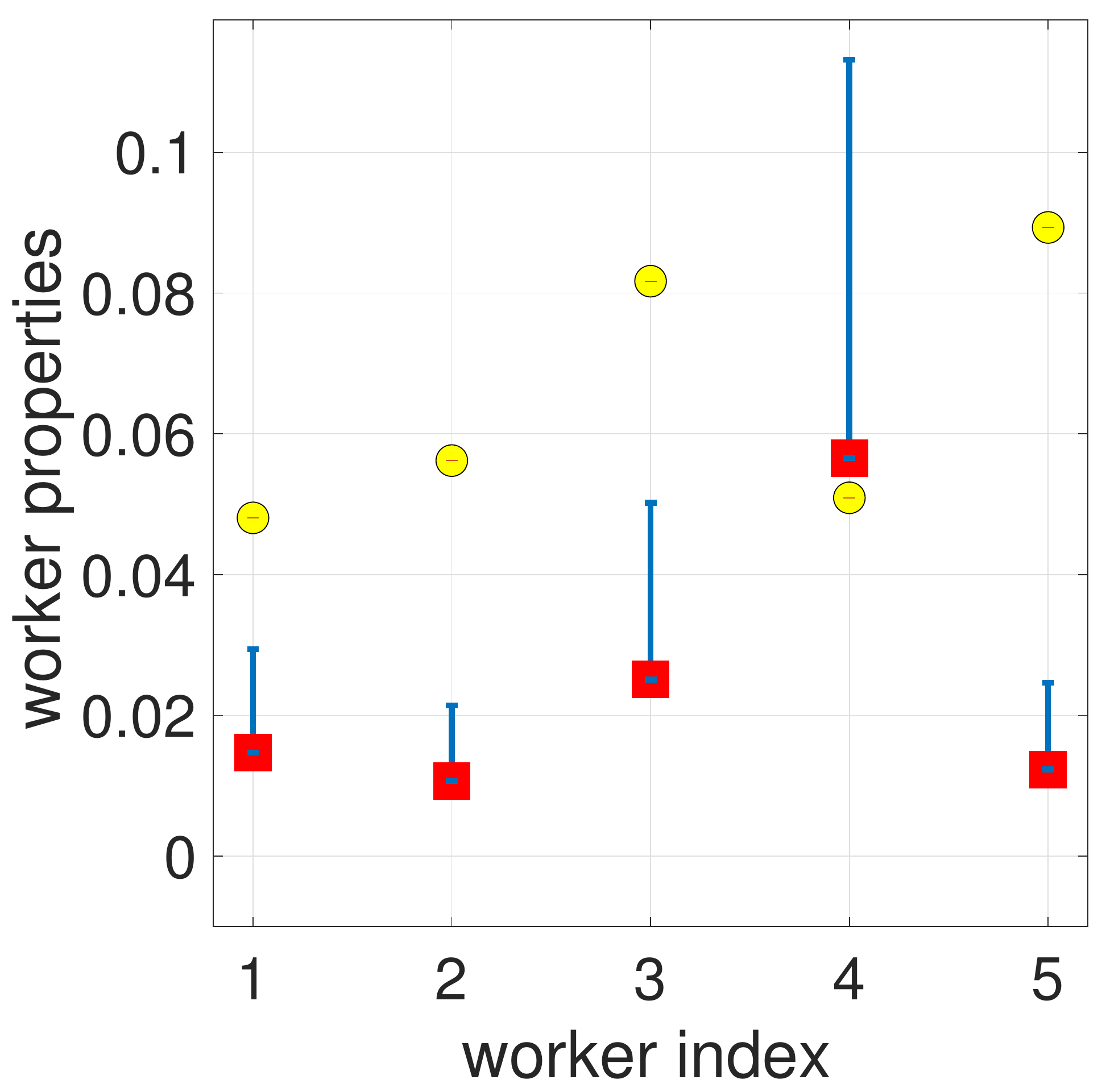} & \includegraphics[width = 0.45\columnwidth]{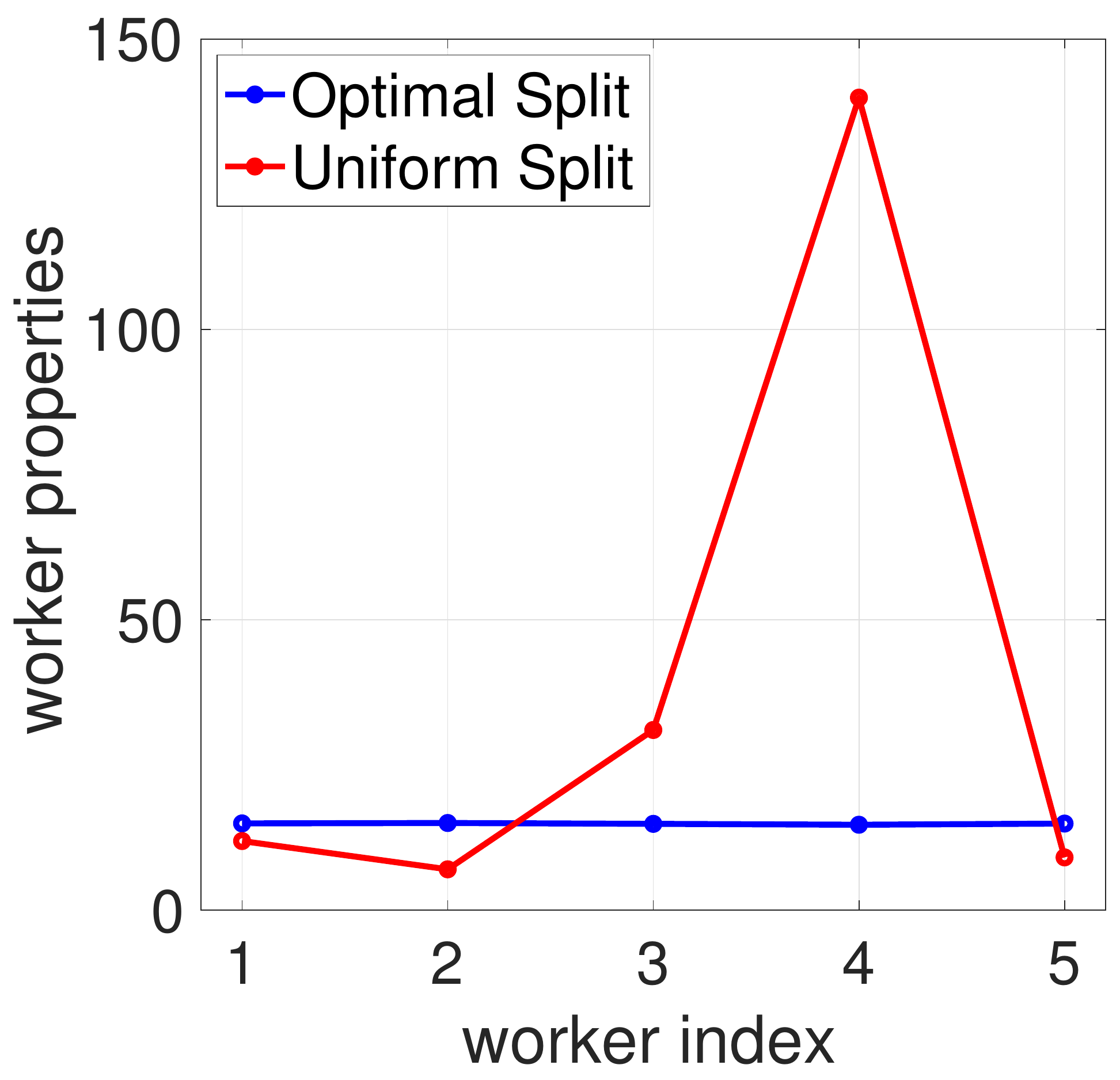} \\
     (a)&(b) 
    \end{tabular}
    \caption{(a) red square and yellow circle show $m_p$ and $c_p$, respectively, and blue bar shows $\sigma_p$; (b) $c_p+\gamma c_p^2+2\gamma\kappa_p c_pm_p+\kappa_p m_p+\kappa_p(\kappa_p-1)\sigma_p^2$.\vspace{-0.2cm}}
    \label{fig:workers_realization1}
\end{figure}

In this subsection, we consider the number of workers is $P=5$, complexity of each task is $C=500$, number of critical tasks is $K=1000$, and redundancy ratio is $\Omega=1.0$ (no redundant tasks). Fig.~\ref{fig:workers_realization1} shows the workers' realization and demonstrations of the two different load split schemes. Fig.~\ref{fig:workers_realization1}~(a) shows the communication delay per iteration (i.e., $c_p$), the average of task computation time (i.e., $m_p=C/\mu_p$), and the standard deviation of task computation time (i.e., $\sigma_p=C/\mu_p$) for each worker $p\in\{1,\dots,5\}$. Fig.~\ref{fig:workers_realization1}~(b) shows $E[T_{p,\kappa_p}]+\gamma E[T_{p,\kappa_p}^2]$ for each worker $p\in\{1,\dots,P\}$, as it is used for identifying the distance between two distributions in this paper, see \eqref{equ:distance}. As we see, this metric is the same for all workers considering our proposed optimal split, while showing a notable variance for the uniform split.

\begin{figure}
    \centering
    \includegraphics[width = 0.9 \columnwidth]{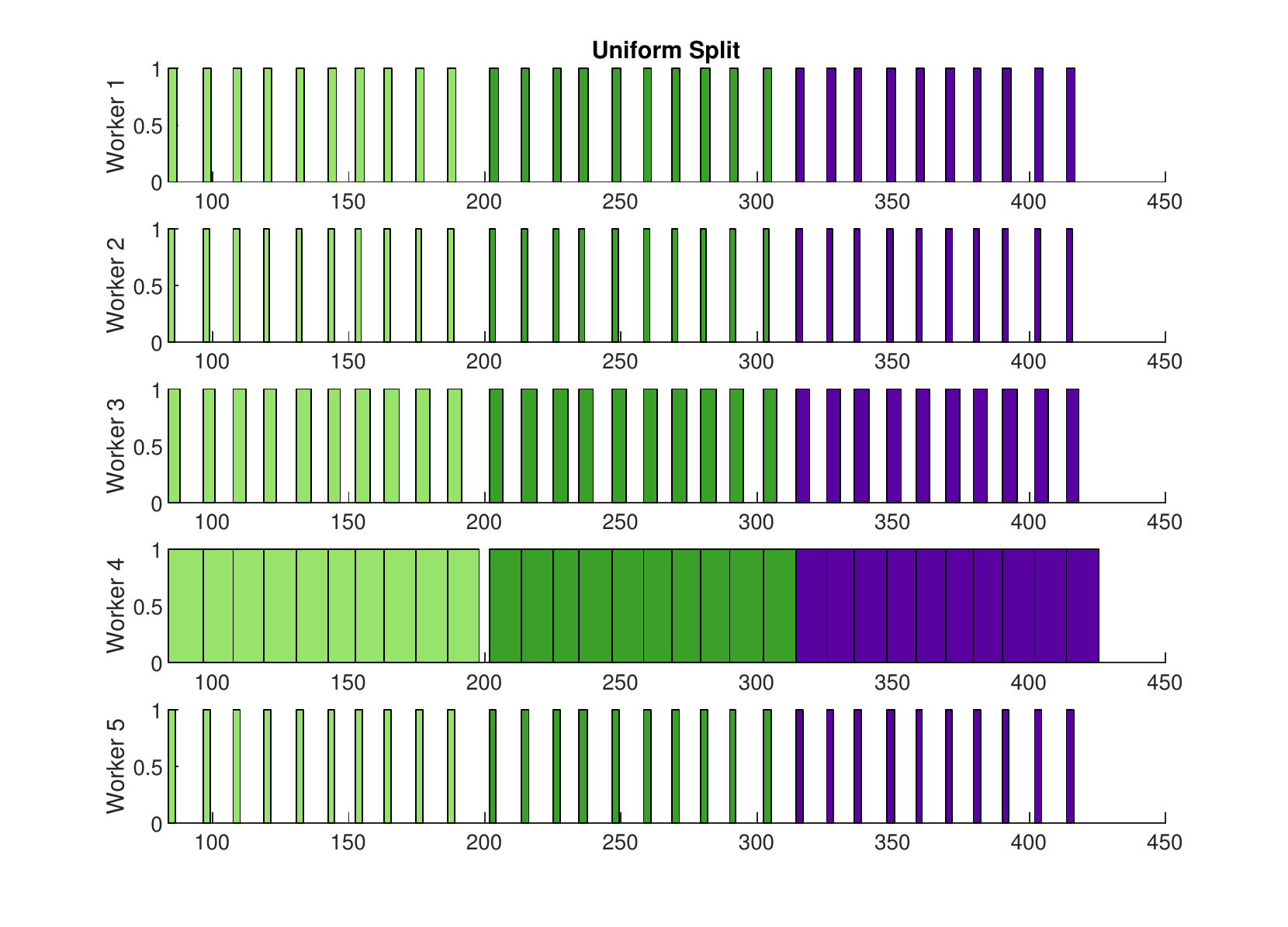}
    \includegraphics[width = 0.9 \columnwidth]{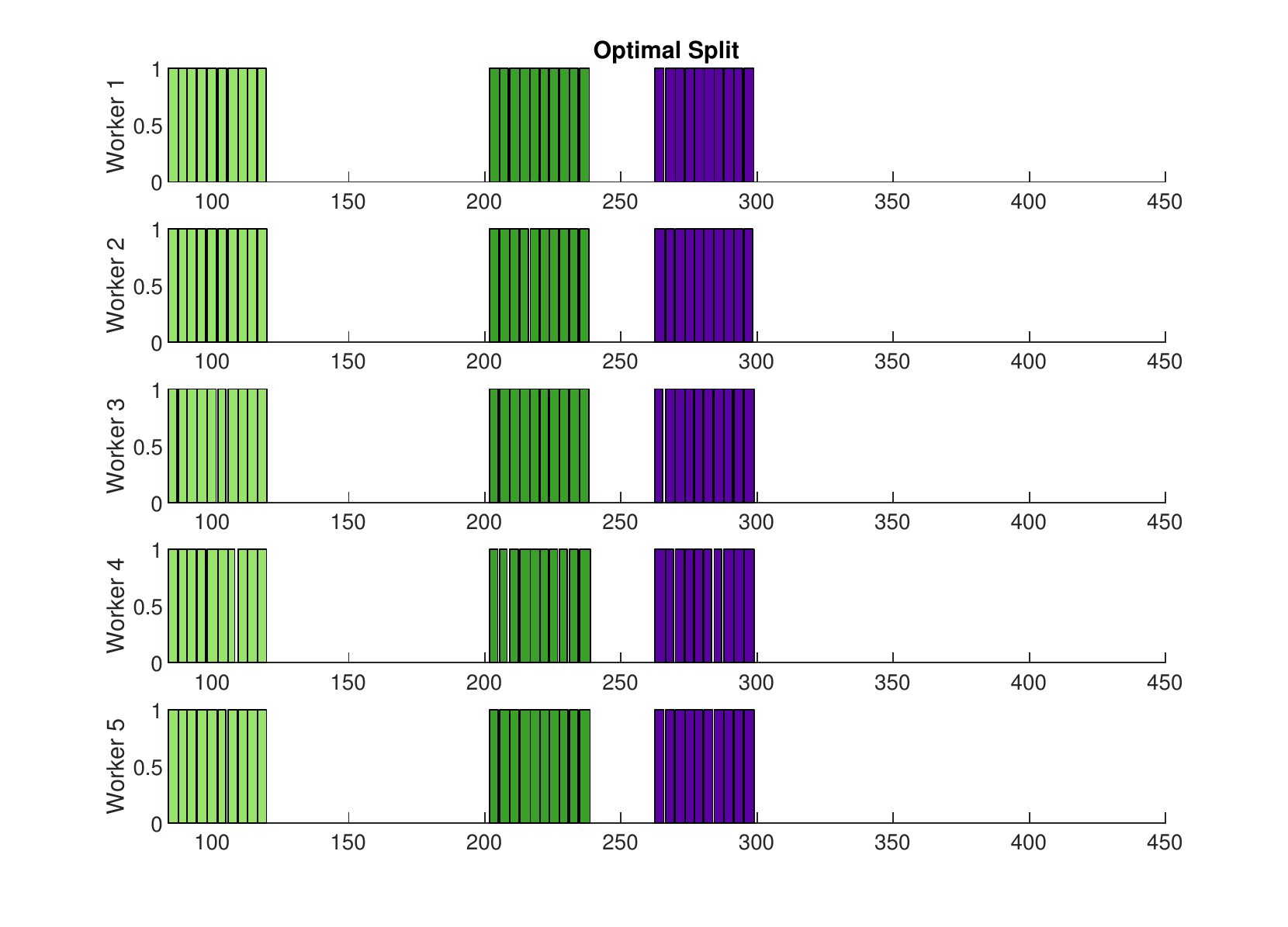}
    \caption{Each job is marked with a specific color, and each colored block shows one iteration of a job. The horizontal axis shows the time and the vertical axis shows if the worker is active or not.\vspace{-0.2cm}}
    \label{fig:streaming_opt_vs_unif}
\end{figure}

For the first $J=3$ arrived jobs according to the Poisson distribution with rate $\lambda=0.01$, we deployed individually the optimal and uniform splits and illustrated their performances over the time in Fig.~\ref{fig:streaming_opt_vs_unif}. This figure shows the state of each worker (being active or idle) over the time for both uniform and optimal solutions, and it shows the time it takes for each worker to respond to an iterative job is notably smaller for the optimal solution compared to the uniform solution. Moreover, the uniform solution experiences a notable waiting time in the queue as the corresponding system is not stable due to the large service time.

\subsection{Execution Delay Analysis}

\begin{figure}
    \centering
    \includegraphics[width = 0.85 \columnwidth]{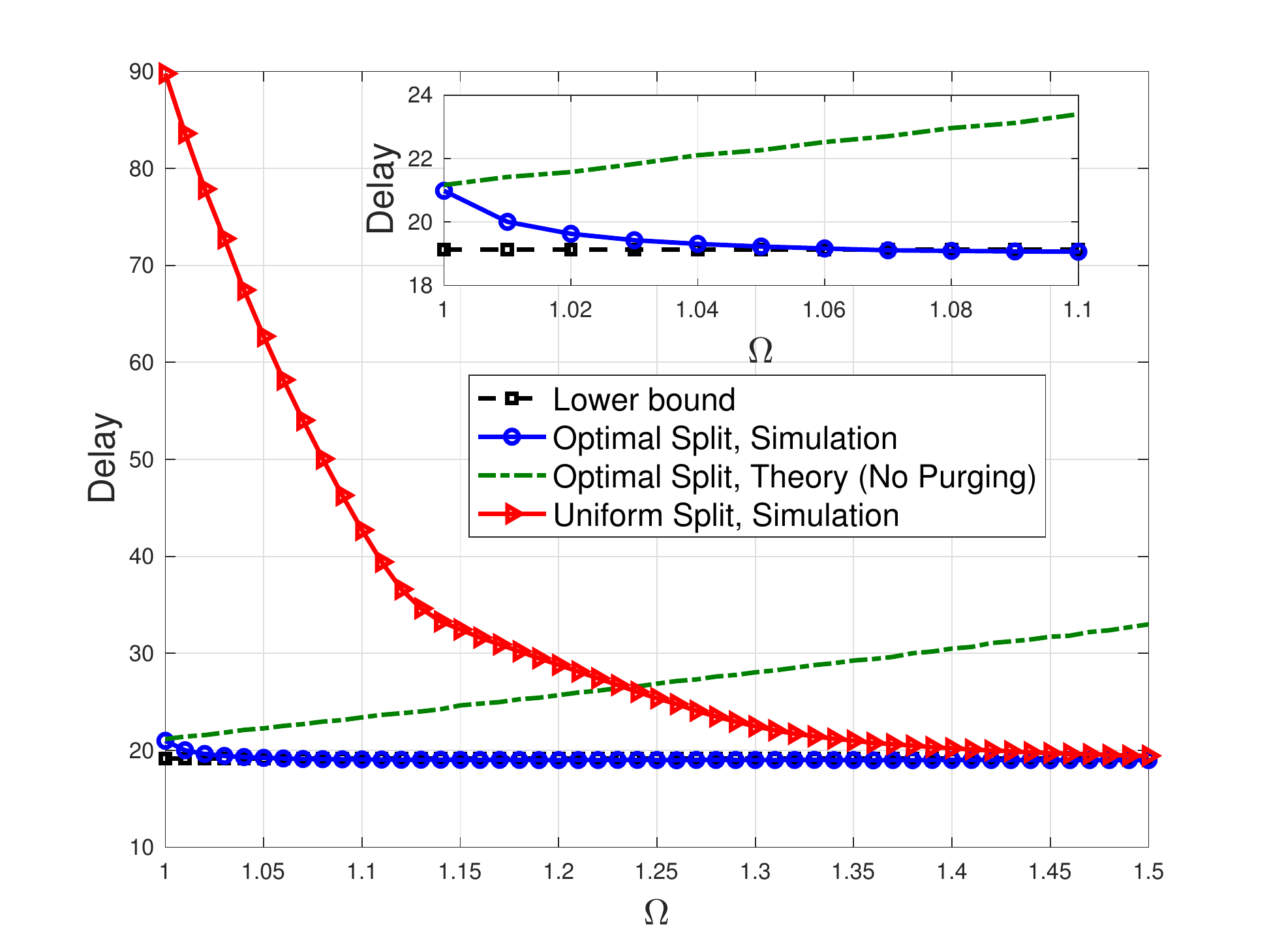}
    \includegraphics[width = 0.85 \columnwidth]{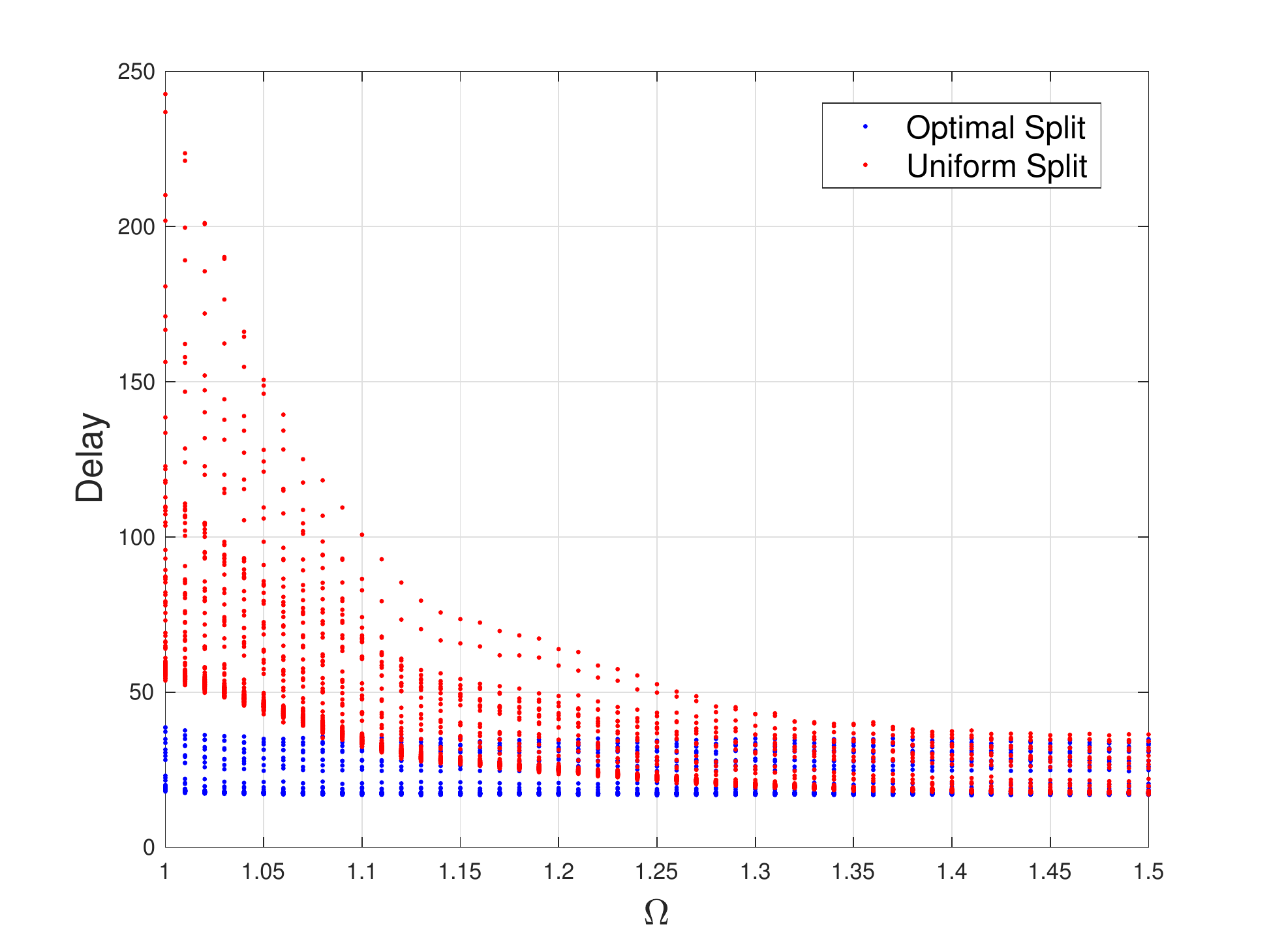}
    \caption{In-order job delay obtained for $J=1000$ and $K=1000$. The upper panel is the average values; and the lower panel is the realizations.}
    \label{fig:job_delay}
\end{figure}

In this subsection, we consider $P=5$, $C=500$, and $K=1000$. We note that in this set of simulations, a stronger set of workers than the previous subsection (workers with higher value of $\mu_p$) are acquired to keep the system stable for all values of $\Omega$ and for both uniform and optimal solutions. However, the workers' realization is not depicted for brevity, The average job execution delay is computed for $J=1000$ jobs arrived at the master node according to the Poisson distribution with parameter $\lambda=0.01$. Fig.~\ref{fig:job_delay}, top panel, shows the the average delay with respect to parameter $\Omega$. The red and blue curves represent the empirical average delay obtained using uniform solution and optimal solution, respectively. The green curve shows the theoretical average delay, which does not take the purging into account, see \eqref{eq:PK}, and the black curve shows the theoretical lower bound, see \eqref{eq:LB}. Fig.~\ref{fig:job_delay}, bottom panel, shows $100$ realizations of the empirical job delay for both uniform and optimal solutions.

As we see, our proposes optimal solution results in a dramatically lower values of the average job execution delay compared to the uniform solution, particularly when the redundancy ratio is low, and it reaches the theoretically-driven lower bound by just adding a small amount of redundant computations, i.e., $\Omega\approx1.06$. It is easy to infer that when there is a large amount of redundant computations in the system, the effect of load split is less important, and thus both uniform and optimal solutions show similar performances. We also note that when there is no purging, the delay increases with increasing $\Omega$, and it justifies why the theoretical value of the delay almost coincides with the simulation results for $\Omega=1$ and it diverges with increasing $\Omega$.

\subsection{Importance of Code Optimization}

\begin{figure}
    \centering
    \begin{tabular}{cc}
     \includegraphics[width = 0.45\columnwidth]{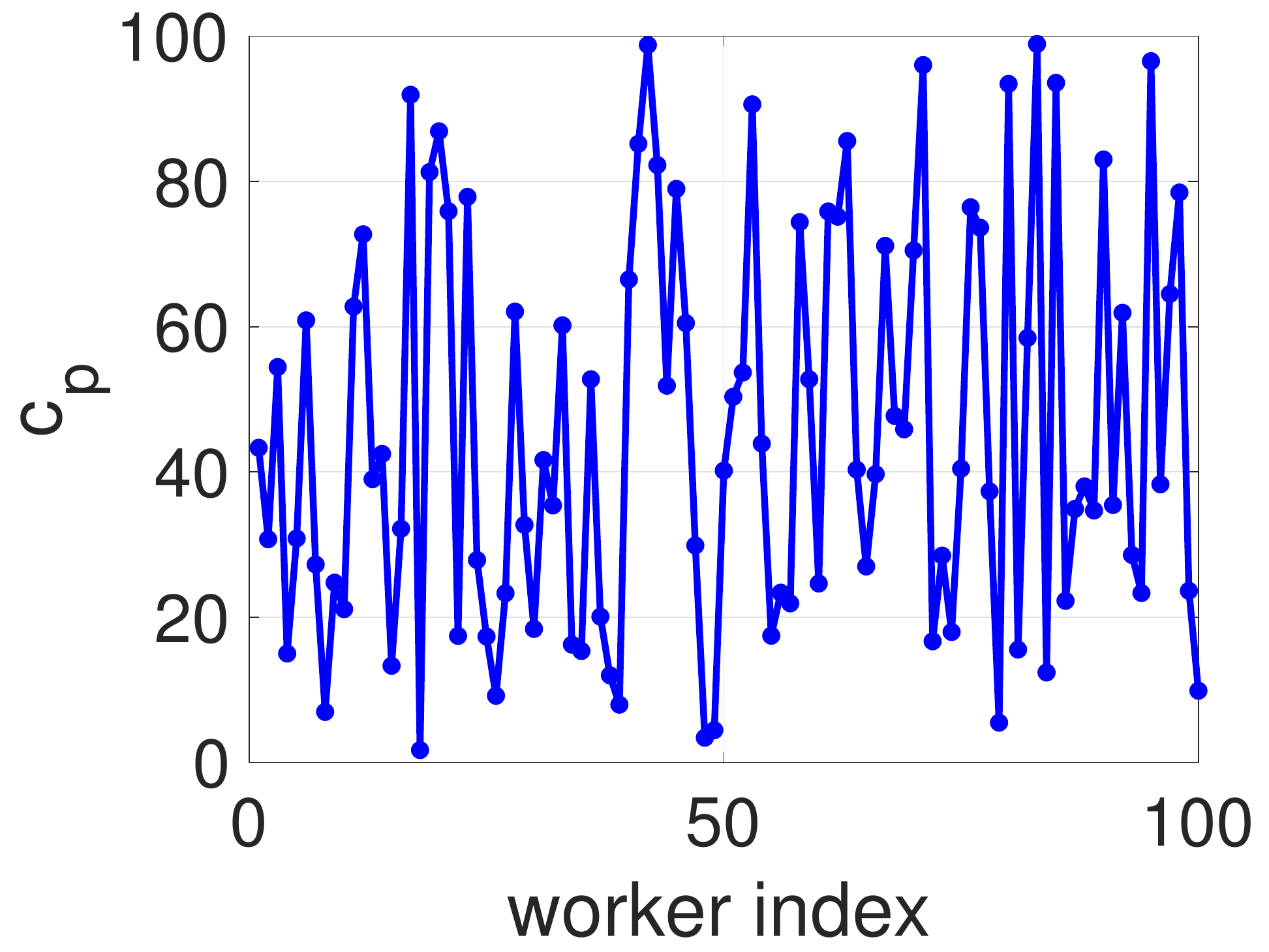} & \includegraphics[width = 0.44\columnwidth]{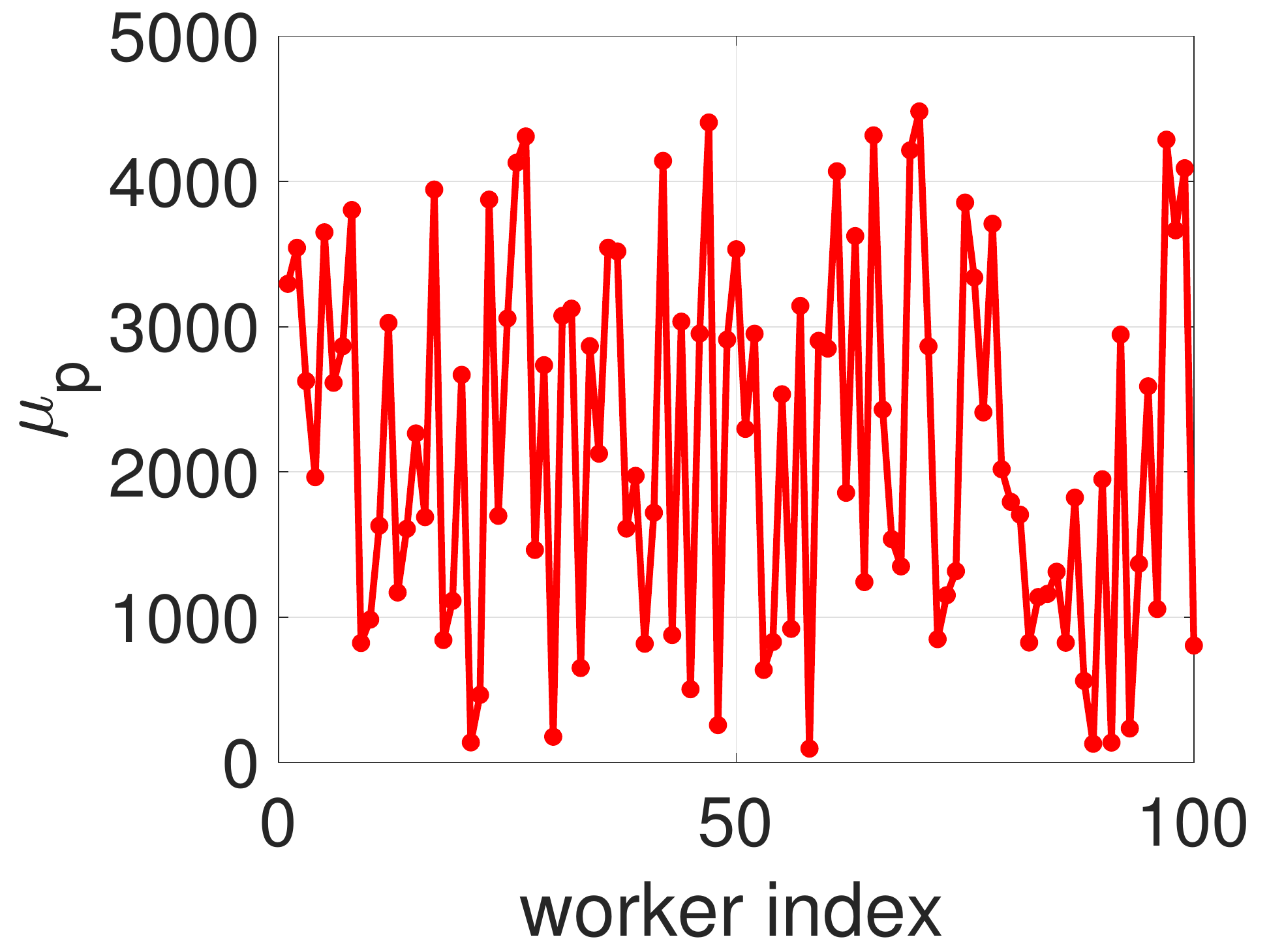}
    \end{tabular}
    \caption{A heterogeneous cluster of $P=100$ workers.\vspace{-0.5cm}}
    \label{fig:workers_realization2}
\end{figure}

In this subsection, we study the affect of code parameters optimization. In this set of simulations, we consider $P=100$ heterogeneous workers, depicted in Fig.~\ref{fig:workers_realization2}. We highlight again that $C$ and $K$ are the code parameters, and we assume $Z\triangleq KC$ is fixed (the relation might be different depending the computational coding scheme). As we elaborated in the paper, for an ideal solution, all workers contribute to the computational results and they all finish their assignment regarding one iteration of a job at the same time. However, in practice due to the communication delay and quantization effects, there is mismatch among distributions of the time it takes for each worker to finish its assignment related to one job iteration, i.e.,  $T_{p,\kappa_p}$ for $p\in\{1,\dots,P\}$.

\begin{figure}
    \centering
     \includegraphics[width = 0.85\columnwidth]{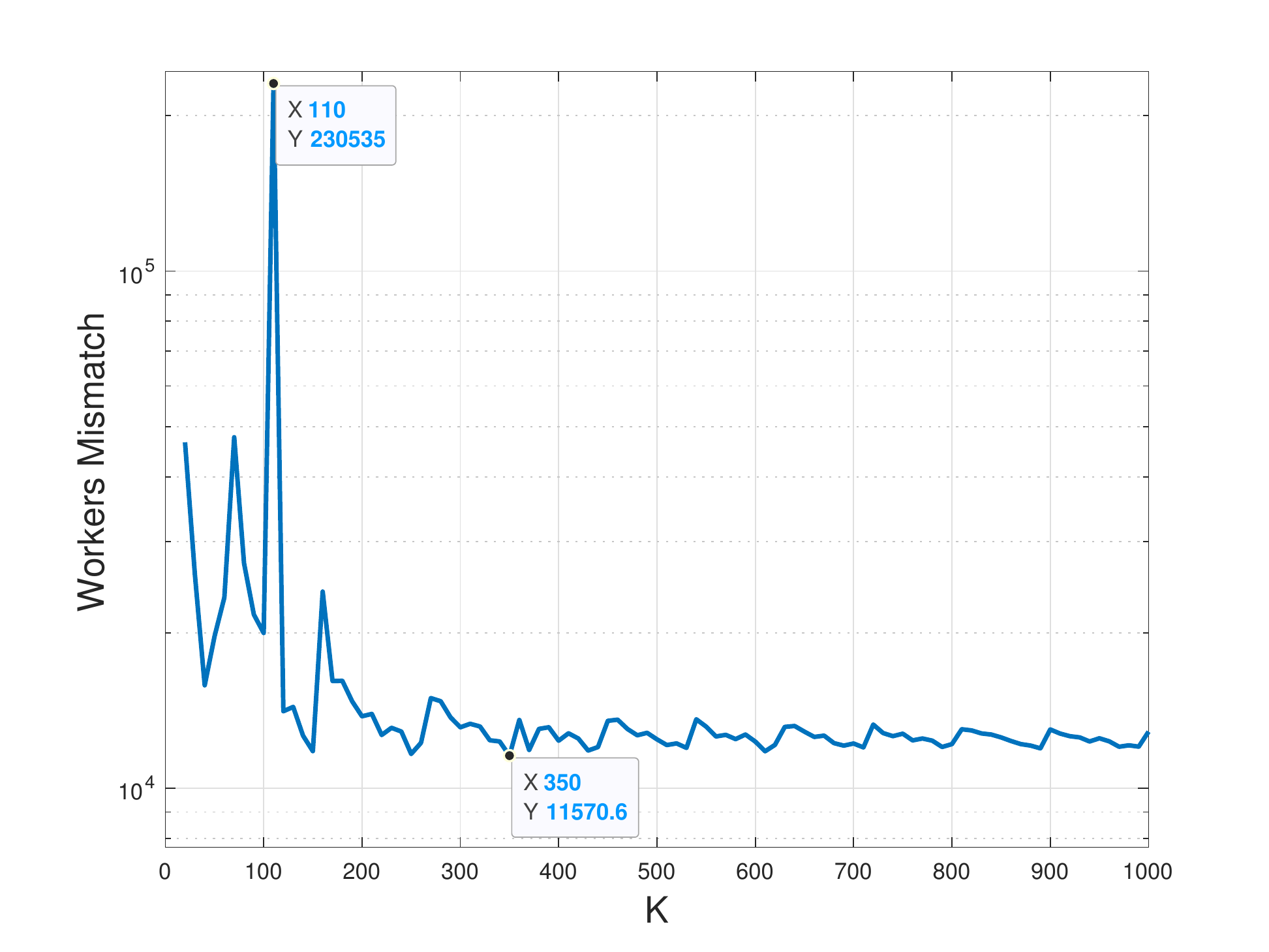}
    \caption{The effect of code parameter $K$ on \textit{mismatch} among the workers.}
    \label{fig:workers_mismatch}
\end{figure}

Thus, we optimize the code parameter $K$, and consequently $C$, such that metric \textit{mismatch}, introduced in \eqref{eq:mismatch}, and corresponding to the optimal split is minimized, as suggested in Algorithm~\ref{alg1}. For this purpose, we have illustrated \textit{mismatch} versus parameter $K$ in Fig.~\ref{fig:workers_mismatch}. As we see the minimum value is obtained when $K=350$, and the relationship between \textit{mismatch} and $K$ is not a strictly decreasing.

\begin{figure}
    \centering
     \includegraphics[width = 0.9\columnwidth]{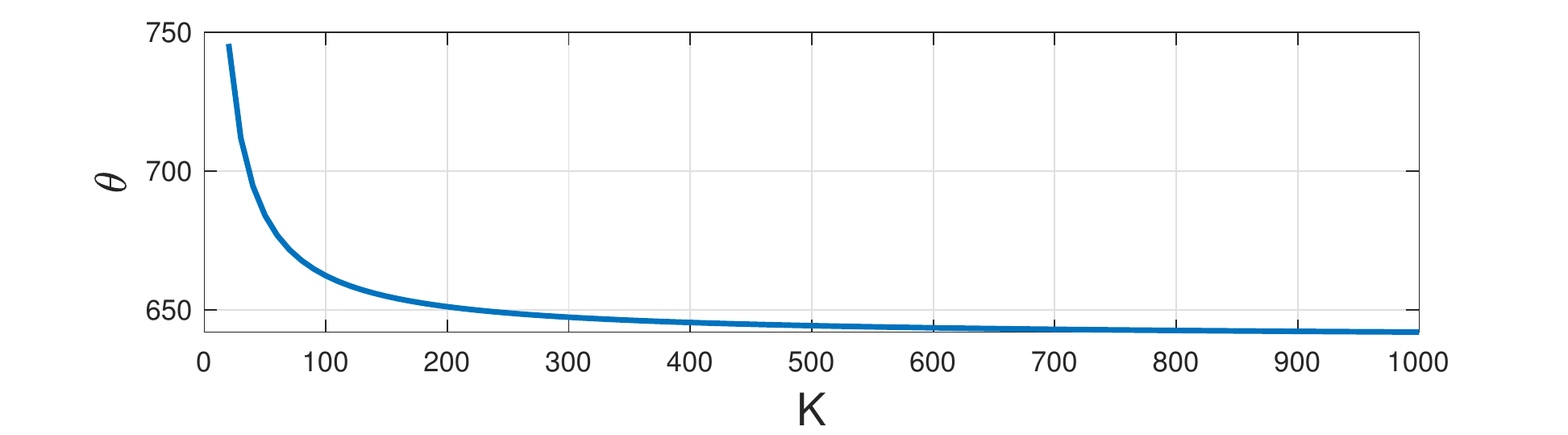}
     \includegraphics[width = 0.9\columnwidth]{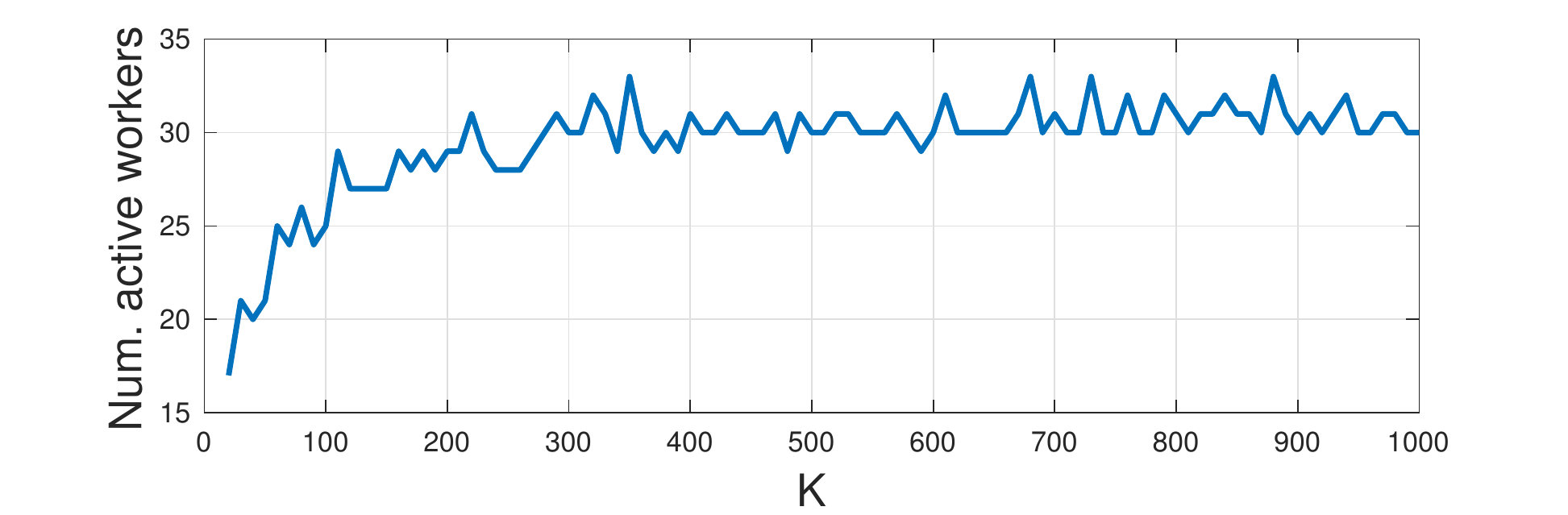}
    \caption{The effect of code parameter $K$ on value of $\theta$ and the number of active workers, for the optimal load split.\vspace{-0.2cm}}
    \label{fig:theta_active}
    \vspace{-0.3cm}
\end{figure}

Finally, Fig.~\ref{fig:theta_active} shows the value of $\theta$ and the number of active workers, obtained via the optimal split, with respect to the code parameter $K$. As we see, for the optimum value of $K$, $\theta=646.24$ and $33$ out of the $100$ workers are active. Finally, Table~\ref{tab:my_label} shows the average job execution delay corresponding to the lowest mismatch ($K=350$), highest mismatch ($K=110$), and two other arbitrary choices. As we observe, the average execution delay is more than $16\%$ lower for the choice of code parameter that results in the least mismatch value compared to the one that results in the highest mismatch value. Besides, after a certain point, increasing the number of tasks (micro-partitioning the computational load) does not offer notable extra benefits in terms of the job execution delay. This result is consistent with the mismatch result depicted in Fig.~\ref{fig:workers_mismatch}, which shows after a certain point the mismatch curve reaches a plateau. We remind that the observations in this section belong to a coding scheme where the relationship between the number of critical tasks and the complexity of each task is $KC=Z$, and the conclusions might change depending the coding scheme principles. \vspace{-0.0cm}

\begin{table}
    \centering
    \caption{Average job execution delay for various values of code parameter $K$.}
    \begin{tabular}{ccccc}
        \hline
         $K$&$110$ &$200$&$350$ &$510$  \\
         \hline
         Average Delay&9.19E4&8.07E4&7.71E4&7.50E4\\
         \hline
    \end{tabular}
    \vspace{-0.5cm}
    \label{tab:my_label}
\end{table}

\section{Conclusion}
In this paper, we proposed a joint scheduling-coding solution for distributed coded computation of iterative jobs over a system that includes a master node and a cluster of heterogeneous workers. The proposed framework can be used in conjunction with many state-of-the-art computational coding schemes to reduce the streaming delay for many iterative data-hungry and intensive computational tasks. An interesting future direction is to modify the scheme for a setting that does not include a master node. Moreover, proposing  
a joint scheduling-coding for stream distributed computation of iterative jobs in a network of heterogeneous interconnected workers is another interesting future research direction. Finally, incorporating the delay of encoding, decoding, iteration updates, and dataset download into the design is a promising future direction. The authors have provided public access to their code and data at \href{https://codeocean.com/capsule/48a9da3f-a31b-4d52-8a84-07c3a3cb374e/}{https://codeocean.com/capsule/48a9da3f-a31b-4d52-8a84-07c3a3cb374e/}.

\section{Acknowledgments}
This work is supported by the European Regional Development Fund (FEDER), through the Regional Operational Programme of Centre (CENTRO 2020) of the Portugal 2020 framework and FCT under the MIT Portugal Program [Project SNOB-5G with Nr. 045929(CENTRO-01-0247-FEDER-045929)].

\section{appendix}

In this section, we review the distributed gradient descent algorithm as an example of iterative algorithms that can be incorporated into our framework. Here, the $j$-th computational job is training a neural network $\mathcal{N}(j)$ using a large dataset $X(j)=\{(\overline{x}_1,y_1),\dots,(\overline{x}_N,y_N)\}$ that consists of $N$ training samples, via stochastic gradient descent (SGD) method\cite{Goodfellow-et-al-2016}. Each tuple $(\overline{x}_i,y_i)$ includes a feature vector $\overline{x}_i\in\mathbb{R}^p$ and an output $y_i\in\mathbb{R}$. The learning job is to solve the following optimization problem,
\begin{equation*}
    W^*(j)=\underset{W}{\text{argmin}}{\sum_{i=1}^{N}Q(W,\overline{x}_i,y_i)},
\end{equation*}
where $W$ determines the weights of the neural network and $Q(.)$ is the fitting loss function. The  gradient-based method aims at solving this optimization problem by iteratively performing the following process,
\begin{align}
    &G=\sum_{(x_i,y_i)\in X(j)}\nabla Q(W^\textit{itr},\overline{x}_i,y_i),\label{equ:gradient}\\
    &W^{\textit{itr}+1}=h\left(W^\textit{itr},G\right),\label{equ:combine}
\end{align}
Equation (\ref{equ:gradient}) is the full-gradient of the model with respect to its weights, and equation (\ref{equ:combine}) is the weight update step where $h(.)$ is a gradient-based optimizer. The distributed gradient descent suggests performing (\ref{equ:gradient}) in a distributed fashion over smaller chunks of the dataset distributed among the workers. The master node receives all the results from the workers, finds the summation to obtain the full-gradient, and performs (\ref{equ:combine}) to update the weights. 

To address the straggler problem, a distributed coded gradient descent was recently proposed \cite{tandon2017gradient,raviv2018gradient} and is reviewed here. The dataset $X(j)$ is partitioned into $m$ disjoint chunks, i.e., $X_1(j)$, \dots, $X_m(j)$. The $r$-th task related to completion of iteration \textit{itr} of $j$-th job is performing the SGD individually on $d$ disjoint chunks of the dataset and sending a linear combination of the results to the master node. The master node uses the results from all workers to identify the summation of all partial gradients. 

To this end, we define a sparse matrix $B$ with size $K\Omega\times m$ such that each row has exactly $d<m$ non-zero elements. Each task corresponds to one row of this matrix, such that the location of non-zero elements determine the index of dataset chunks and the value of non-zero elements determine the combination rule. At each iteration, the master node sends the current weights of the model and requests a subset of the tasks from each worker (via our optimal split solution). Each worker performs the assigned tasks sequentially and sends the results back to the master node as soon as a task is finished. The master node is capable of obtaining the full gradient update from any $K$ task results if the full-one row is in the span of any submatrix that consists of $K$ rows of matrix $B$. Then, it completes the iteration by performing the weight update (\ref{equ:combine}). Several constructions have been devised in the literature for matrix $B$, and correspondingly the values for $K$, $\Omega$, $m$, and $d$, e.g., \cite{dutta2019short,tandon2017gradient,raviv2018gradient,halbawi2018improving,raviv2020gradient}.

\begin{figure}
    \centering
    \includegraphics[width = 0.85 \columnwidth]{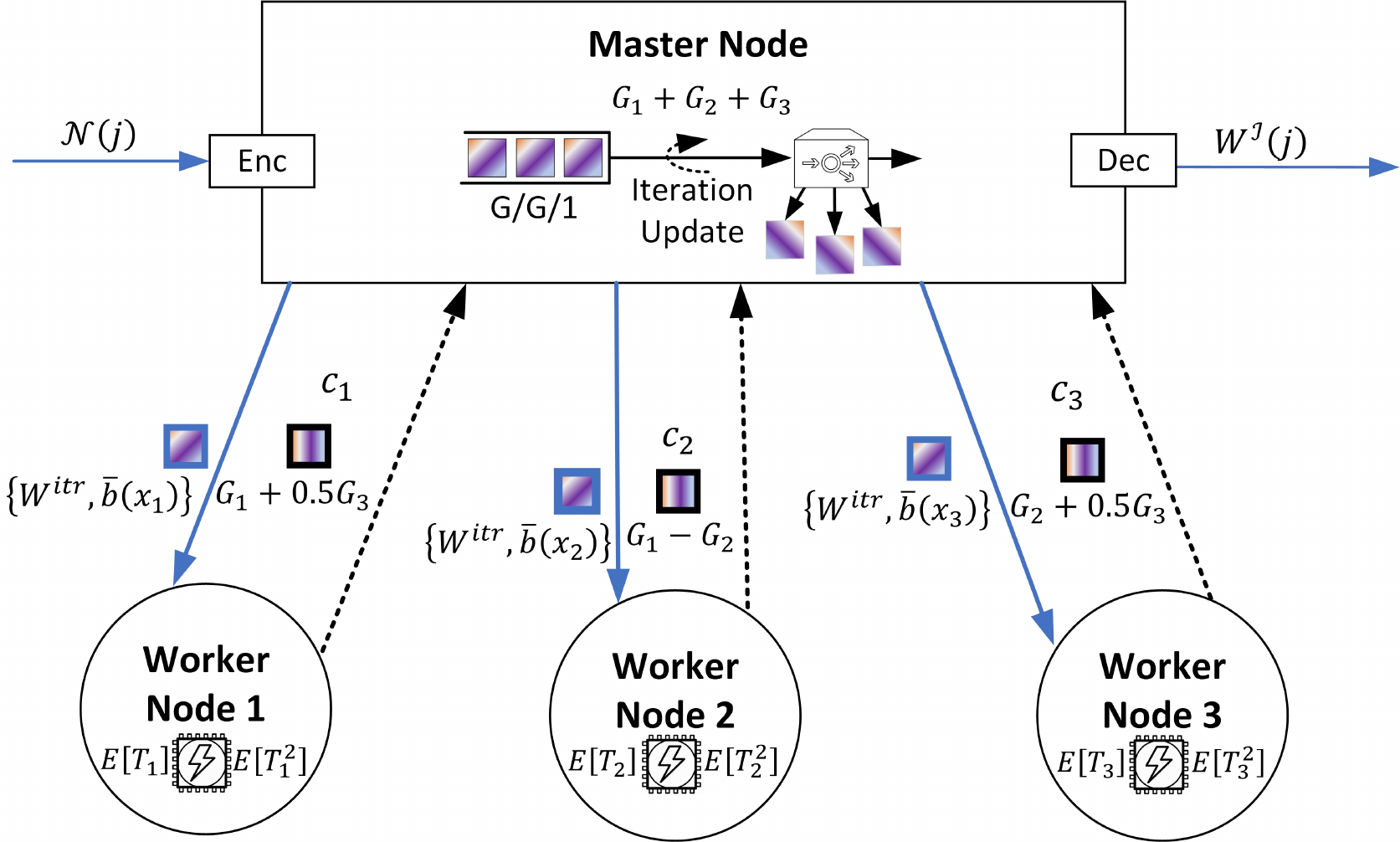}
    \caption{Coded distributed Gradient descent computation, example for $K=2$, $\Omega=1.5$, $m=3$, and $d=2$.\vspace{-0.2cm}}
    \label{fig:GD_example}
    \vspace{-0.3cm}
\end{figure} 

\begin{example}
Consider an example where $K=2$, $\Omega=1.5$, $m=3$, and $d=2$. The following choice for matrix $\mathbf{B}$:
$$\mathbf{B}=\left(
\begin{array}{ccc}
     1&0&0.5  \\
     1&-1&0   \\
     0&1&0.5
\end{array}
\right),$$
enables the final gradient descent update to be obtained from the result of any $K=2$ successful task results, see Fig.~\ref{fig:GD_example}. More specifically, the dataset is partitioned into $m=3$ chunks. There are $K\Omega=3$ tasks: The first task is computing the gradient updates on the first and third partitions as $G_1$ and $G_3$, and sending $G_1+0.5G_3$ back to the master node. The second task is computing the gradient updates on first and second partitions as $G_1$ and $G_2$, and sending $G_1-G_2$ back to the master node. The third task is computing the gradient updates on second and third partitions as $G_2$ and $G_3$, and sending $G_2+0.5G_3$ back to the master node. As it is apparent, the value of $G_1+G_2+G_3$ can be obtained from any $K=2$ successful task results.
\end{example}



\clearpage

\bibliographystyle{IEEEtran}
\bibliography{references}

\begin{thebibliography}{10}
\providecommand{\url}[1]{#1}
\csname url@samestyle\endcsname
\providecommand{\newblock}{\relax}
\providecommand{\bibinfo}[2]{#2}
\providecommand{\BIBentrySTDinterwordspacing}{\spaceskip=0pt\relax}
\providecommand{\BIBentryALTinterwordstretchfactor}{4}
\providecommand{\BIBentryALTinterwordspacing}{\spaceskip=\fontdimen2\font plus
\BIBentryALTinterwordstretchfactor\fontdimen3\font minus
  \fontdimen4\font\relax}
\providecommand{\BIBforeignlanguage}[2]{{%
\expandafter\ifx\csname l@#1\endcsname\relax
\typeout{** WARNING: IEEEtran.bst: No hyphenation pattern has been}%
\typeout{** loaded for the language `#1'. Using the pattern for}%
\typeout{** the default language instead.}%
\else
\language=\csname l@#1\endcsname
\fi
#2}}
\providecommand{\BIBdecl}{\relax}
\BIBdecl

\bibitem{XINXingML}
\BIBentryALTinterwordspacing
E.~P. Xing, Q.~Ho, P.~Xie, and D.~Wei, ``Strategies and principles of
  distributed machine learning on big data,'' \emph{Engineering}, vol.~2,
  no.~2, pp. 179--195, 2016. [Online]. Available:
  \url{https://www.sciencedirect.com/science/article/pii/S2095809916309468}
\BIBentrySTDinterwordspacing

\bibitem{dean2013tail}
J.~Dean and L.~A. Barroso, ``The tail at scale,'' \emph{Communications of the
  ACM}, vol.~56, no.~2, pp. 74--80, 2013.

\bibitem{joshi2017efficient}
G.~Joshi, E.~Soljanin, and G.~Wornell, ``Efficient redundancy techniques for
  latency reduction in cloud systems,'' \emph{ACM Transactions on Modeling and
  Performance Evaluation of Computing Systems (TOMPECS)}, vol.~2, no.~2, p.~12,
  2017.

\bibitem{Reisizadeh}
A.~Reisizadeh, S.~Prakash, R.~Pedarsani, and A.~S. Avestimehr, ``Coded
  computation over heterogeneous clusters,'' \emph{IEEE Transactions on
  Information Theory}, vol.~65, no.~7, pp. 4227--4242, 2019.

\bibitem{Avestimehr}
C.-S. Yang, R.~Pedarsani, and A.~S. Avestimehr, ``Timely coded computing,'' in
  \emph{{IEEE} International Symposium on Information Theory (ISIT)}, 2019, pp.
  2798--2802.

\bibitem{sheth2018application}
U.~Sheth, S.~Dutta, M.~Chaudhari, H.~Jeong, Y.~Yang, J.~Kohonen, T.~Roos, and
  P.~Grover, ``An application of storage-optimal matdot codes for coded matrix
  multiplication: Fast k-nearest neighbors estimation,'' in \emph{2018 IEEE
  International Conference on Big Data (Big Data)}.\hskip 1em plus 0.5em minus
  0.4em\relax IEEE, 2018, pp. 1113--1120.

\bibitem{d2020gasp}
R.~G. D’Oliveira, S.~El~Rouayheb, and D.~Karpuk, ``{GASP} codes for secure
  distributed matrix multiplication,'' \emph{IEEE Transactions on Information
  Theory}, vol.~66, no.~7, pp. 4038--4050, 2020.

\bibitem{wahabfederated}
O.~A. Wahab, A.~Mourad, H.~Otrok, and T.~Taleb, ``Federated machine learning:
  Survey, multi-level classification, desirable criteria and future directions
  in communication and networking systems.''

\bibitem{cohen2021stream}
A.~Cohen, G.~Thiran, H.~Esfahanizadeh, and M.~M{\'e}dard, ``Stream distributed
  coded computing,'' \emph{IEEE Journal on Selected Areas in Information
  Theory}, vol.~2, no.~3, pp. 1025--1040, 2021.

\bibitem{yu2017polynomialn}
Q.~Yu, M.~A. Maddah-Ali, and S.~Avestimehr, ``Polynomial codes: an optimal
  design for high-dimensional coded matrix multiplication,'' in \emph{NIPS},
  2017.

\bibitem{raviv2020gradient}
N.~Raviv, I.~Tamo, R.~Tandon, and A.~G. Dimakis, ``Gradient coding from cyclic
  {MDS} codes and expander graphs,'' \emph{IEEE Transactions on Information
  Theory}, vol.~66, no.~12, pp. 7475--7489, 2020.

\bibitem{karakus2019redundancy}
C.~Karakus, Y.~Sun, S.~Diggavi, and W.~Yin, ``Redundancy techniques for
  straggler mitigation in distributed optimization and learning,'' \emph{The
  Journal of Machine Learning Research}, vol.~20, no.~1, pp. 2619--2665, 2019.

\bibitem{dutta2019short}
S.~Dutta, V.~Cadambe, and P.~Grover, ``“{S}hort-{D}ot”: Computing large
  linear transforms distributedly using coded short dot products,'' \emph{IEEE
  Transactions on Information Theory}, vol.~65, no.~10, pp. 6171--6193, 2019.

\bibitem{tandon2017gradient}
R.~Tandon, Q.~Lei, A.~G. Dimakis, and N.~Karampatziakis, ``Gradient coding:
  Avoiding stragglers in distributed learning,'' in \emph{International
  Conference on Machine Learning}.\hskip 1em plus 0.5em minus 0.4em\relax PMLR,
  2017, pp. 3368--3376.

\bibitem{halbawi2018improving}
W.~Halbawi, N.~Azizan, F.~Salehi, and B.~Hassibi, ``Improving distributed
  gradient descent using reed-solomon codes,'' in \emph{2018 IEEE International
  Symposium on Information Theory (ISIT)}.\hskip 1em plus 0.5em minus
  0.4em\relax IEEE, 2018, pp. 2027--2031.

\bibitem{maity2019robust}
R.~K. Maity, A.~S. Rawa, and A.~Mazumdar, ``Robust gradient descent via moment
  encoding and ldpc codes,'' in \emph{2019 IEEE International Symposium on
  Information Theory (ISIT)}.\hskip 1em plus 0.5em minus 0.4em\relax IEEE,
  2019, pp. 2734--2738.

\bibitem{kosaian2018learning}
J.~Kosaian, K.~Rashmi, and S.~Venkataraman, ``Learning a code: Machine learning
  for approximate non-linear coded computation,'' \emph{arXiv preprint
  arXiv:1806.01259}, 2018.

\bibitem{8002642}
K.~Lee, M.~Lam, R.~Pedarsani, D.~Papailiopoulos, and K.~Ramchandran, ``Speeding
  up distributed machine learning using codes,'' \emph{IEEE Transactions on
  Information Theory}, vol.~64, no.~3, pp. 1514--1529, 2018.

\bibitem{8437852}
S.~Dutta, Z.~Bai, H.~Jeong, T.~M. Low, and P.~Grover, ``A unified coded deep
  neural network training strategy based on generalized polydot codes,'' in
  \emph{2018 IEEE International Symposium on Information Theory (ISIT)}, 2018,
  pp. 1585--1589.

\bibitem{lee2017high}
K.~Lee, C.~Suh, and K.~Ramchandran, ``High-dimensional coded matrix
  multiplication,'' in \emph{2017 IEEE International Symposium on Information
  Theory (ISIT)}.\hskip 1em plus 0.5em minus 0.4em\relax IEEE, 2017, pp.
  2418--2422.

\bibitem{suh2017matrix}
G.~Suh, K.~Lee, and C.~Suh, ``Matrix sparsification for coded matrix
  multiplication,'' in \emph{2017 55th Annual Allerton Conference on
  Communication, Control, and Computing (Allerton)}.\hskip 1em plus 0.5em minus
  0.4em\relax IEEE, 2017, pp. 1271--1278.

\bibitem{baharav2018straggler}
T.~Baharav, K.~Lee, O.~Ocal, and K.~Ramchandran, ``Straggler-proofing
  massive-scale distributed matrix multiplication with d-dimensional product
  codes,'' in \emph{2018 IEEE International Symposium on Information Theory
  (ISIT)}.\hskip 1em plus 0.5em minus 0.4em\relax IEEE, 2018, pp. 1993--1997.

\bibitem{Dutta2020}
S.~{Dutta}, M.~{Fahim}, F.~{Haddadpour}, H.~{Jeong}, V.~{Cadambe}, and
  P.~{Grover}, ``On the optimal recovery threshold of coded matrix
  multiplication,'' \emph{IEEE Transactions on Information Theory}, vol.~66,
  no.~1, pp. 278--301, 2020.

\bibitem{mallick2019fast}
A.~Mallick, M.~Chaudhari, and G.~Joshi, ``Fast and efficient distributed
  matrix-vector multiplication using rateless fountain codes,'' in \emph{ICASSP
  2019-2019 IEEE International Conference on Acoustics, Speech and Signal
  Processing (ICASSP)}.\hskip 1em plus 0.5em minus 0.4em\relax IEEE, 2019, pp.
  8192--8196.

\bibitem{Maddah-Ali2018CommComp}
S.~Li, M.~A. Maddah-Ali, and A.~S. Avestimehr, ``Fundamental tradeoff between
  computation and communication in distributed computing,'' in \emph{2016 IEEE
  International Symposium on Information Theory (ISIT)}, 2016, pp. 1814--1818.

\bibitem{ye2018communication}
M.~Ye and E.~Abbe, ``Communication-computation efficient gradient coding,'' in
  \emph{International Conference on Machine Learning}.\hskip 1em plus 0.5em
  minus 0.4em\relax PMLR, 2018, pp. 5610--5619.

\bibitem{song2019pliable}
L.~Song, C.~Fragouli, and T.~Zhao, ``A pliable index coding approach to data
  shuffling,'' \emph{IEEE Transactions on Information Theory}, vol.~66, no.~3,
  pp. 1333--1353, 2019.

\bibitem{attia2019near}
M.~A. Attia and R.~Tandon, ``Near optimal coded data shuffling for distributed
  learning,'' \emph{IEEE Transactions on Information Theory}, vol.~65, no.~11,
  pp. 7325--7349, 2019.

\bibitem{dutta2017coded}
S.~Dutta, V.~Cadambe, and P.~Grover, ``Coded convolution for parallel and
  distributed computing within a deadline,'' in \emph{2017 IEEE International
  Symposium on Information Theory (ISIT)}.\hskip 1em plus 0.5em minus
  0.4em\relax IEEE, 2017, pp. 2403--2407.

\bibitem{yang2016fault}
Y.~Yang, P.~Grover, and S.~Kar, ``Fault-tolerant parallel linear filtering
  using compressive sensing,'' in \emph{2016 9th International Symposium on
  Turbo Codes and Iterative Information Processing (ISTC)}.\hskip 1em plus
  0.5em minus 0.4em\relax IEEE, 2016, pp. 201--205.

\bibitem{jeong2018masterless}
H.~Jeong, T.~M. Low, and P.~Grover, ``Masterless coded computing: A
  fully-distributed coded fft algorithm,'' in \emph{2018 56th Annual Allerton
  Conference on Communication, Control, and Computing (Allerton)}.\hskip 1em
  plus 0.5em minus 0.4em\relax IEEE, 2018, pp. 887--894.

\bibitem{yu2017coded}
Q.~Yu, M.~A. Maddah-Ali, and A.~S. Avestimehr, ``Coded fourier transform,'' in
  \emph{2017 55th Annual Allerton Conference on Communication, Control, and
  Computing (Allerton)}.\hskip 1em plus 0.5em minus 0.4em\relax IEEE, 2017, pp.
  494--501.

\bibitem{Klein1975}
L.~Kleinrock, \emph{Queuing Systems Vol. I: Theory}.\hskip 1em plus 0.5em minus
  0.4em\relax New York: Wiley, 1975.

\bibitem{mackay_2019}
D.~J.~C. MacKay, \emph{Information theory, inference, and learning
  algorithms}.\hskip 1em plus 0.5em minus 0.4em\relax Cambridge University
  Press, 2019.

\bibitem{marchal1976approximate}
W.~G. Marchal, ``An approximate formula for waiting time in single server
  queues,'' \emph{AIIE transactions}, vol.~8, no.~4, pp. 473--474, 1976.

\bibitem{gallager2013stochastic}
R.~G. Gallager, \emph{Stochastic processes: theory for applications}.\hskip 1em
  plus 0.5em minus 0.4em\relax Cambridge University Press, 2013.

\bibitem{raviv2018gradient}
N.~Raviv, R.~Tandon, A.~Dimakis, and I.~Tamo, ``Gradient coding from cyclic
  {MDS} codes and expander graphs,'' in \emph{International Conference on
  Machine Learning}.\hskip 1em plus 0.5em minus 0.4em\relax PMLR, 2018, pp.
  4305--4313.

\bibitem{lee2017speeding}
K.~Lee, M.~Lam, R.~Pedarsani, D.~Papailiopoulos, and K.~Ramchandran, ``Speeding
  up distributed machine learning using codes,'' \emph{IEEE Transactions on
  Information Theory}, vol.~64, no.~3, pp. 1514--1529, 2017.

\bibitem{Goodfellow-et-al-2016}
I.~Goodfellow, Y.~Bengio, and A.~Courville, \emph{Deep Learning}.\hskip 1em
  plus 0.5em minus 0.4em\relax MIT Press, 2016,
  \url{http://www.deeplearningbook.org}.

\end{thebibliography}

\end{document}